\documentclass[12pt,rmp,preprintnumbers,floatfix,showpacs,showkeys,tightenlines,nofootinbib]{revtex4}

\usepackage{url}
\usepackage{natbib}
\usepackage{amsmath}
\usepackage{bm}		
\usepackage{graphicx}

\def\program#1{{\sc #1}}
\def\defn#1{``#1''}

\def\ie{i.e.\hbox{}}

\def\Cplusplus{\hbox{C\raise.2ex\hbox{\footnotesize ++}}}
\def\P#1{\phantom{#1}}
\def\jtitembreak{\mbox{}\\}	

\def\csmash#1{\hbox to 0em{\hss{#1}\hss}}
\def\xlsy#1#2{{\setbox0=\hbox{#2}\hbox to \wd0{{#1}\hss}}}
\def\xcsy#1#2{{\setbox0=\hbox{#2}\hbox to \wd0{\hss{#1}\hss}}}

\def\abs{{\text{abs}}}
\def\background{{\text{background}}}
\def\constant{{\text{constant}}}

\def\iperp{{\text{\tt iperp}}}
\def\ipar{{\text{\tt ipar}}}
\def\absexact{{\text{abs,exact}}}
\def\nominalmin{{\text{nominal,min}}}
\def\nominalmax{{\text{nominal,max}}}

\def\ddfrac#1#2{\frac{\displaystyle #1}{\displaystyle #2}}

\def\jtbold{\bf\mathversion{bold}}

\def\aphi{\varphi}		
\def\cphi{\phi}			
\def\wr{w}			

\def\gz{{\sf gz}}
\def\gzw{{\sf w}}
\def\p{{\sf p}}
\def\q{{\sf q}}
\def\R{{\sf R}}

\def\degree{\circ}
\def\del{\nabla}
\def\ltsim{\lesssim}
\def\gtsim{\gtrsim}

\def\tfrac#1#2{{\textstyle\frac{#1}{#2}}}

\begin{document}
\title{Black Hole Excision with Multiple Grid Patches}
\author{\firstname{Jonathan} \surname{Thornburg}}
\email{jthorn@aei.mpg.de}
\homepage{http://www.aei.mpg.de/~jthorn}
\affiliation{Max-Planck-Institut f\"ur Gravitationsphysik,
	     Albert-Einstein-Institut,
	     Am M\"uhlenberg~1, D-14476 Golm, Germany}
\preprint{AEI-2004-032}
%
%
\date{$ $Id: multipatch.tex,v 1.294 2004/07/16 13:40:30 jonathan Exp $ $}


\begin{abstract}
\renewcommand{\thempfootnote}{\fnsymbol{mpfootnote}}
When using black hole excision to numerically evolve a black hole
spacetime with no continuous symmetries, most $3+1$ finite differencing
codes use a Cartesian grid.  It's difficult to do excision on such a
grid because the natural $r = \text{constant}$ excision surface must
be approximated either by a very different shape such as a contained
cube, or by an irregular and non-smooth \defn{LEGO\footnotemark[2] sphere}
which may introduce numerical instabilities into the evolution.
In this paper I describe an alternate scheme, which uses multiple
$\{ r \times (\text{angular coordinates}) \}$ grid patches, each patch
using a different (nonsingular) choice of angular coordinates.
This allows excision on a smooth $r = \text{constant}$ 2-sphere.

I discuss the key design choices in such a multiple-patch scheme,
including the choice of ghost-zone versus internal-boundary treatment
of the interpatch boundaries (I use a ghost-zone scheme), the number
and shape of the patches (I use a 6-patch \defn{inflated-cube} scheme),
the details of how the ghost zones are \defn{synchronized} by
interpolation from neighboring patches, the tensor basis for the
Einstein equations in each patch, and the handling of non-tensor
field variables such as the BSSN $\tilde{\Gamma}^i$
(I use a scheme which requires ghost zones which are twice as wide
for the BSSN conformal factor $\cphi$ as for $\tilde{\Gamma}^i$ and
the other BSSN field variables).

I present sample numerical results from a prototype implementation
of this scheme.  This code simulates the time evolution of the
(asymptotically flat) spacetime around a single (excised) black hole,
using 4th~order finite differencing in space and time.  Using Kerr
initial data with $J/m^2 = 0.6$, I present evolutions to $t \gtsim 1500m$.
The lifetime of these evolutions appears to be limited only by
outer boundary instabilities, not by any excision instabilities
or by any problems inherent to the multiple-patch scheme.
\end{abstract}


\maketitle

\begingroup
\renewcommand{\thefootnote}{\fnsymbol{footnote}}
\footnotetext[2]{
		LEGO is a registered trademark of the LEGO Company,
		\url{http://www.lego.com}.
		}
\endgroup


\section{Introduction}
\label{sect-introduction}

When time-evolving a black hole spacetime in $3+1$ (Cauchy) numerical
relativity, the numerical computation must somehow be kept from
encountering the singularity(ies) within the black hole(s).  There
are three common means of doing this:
freezing slicings (\citet{Lichnerowicz44,Smarr78b}),
the Brandt-Br\"{u}gmann ``puncture'' method (\citet{Brandt97b}),
and excision
(\citet{Unruh84,Thornburg85,Thornburg87,Seidel92a,Thornburg93,Anninos94e}).
In this paper I focus on excision in the context of finite-difference
evolutions, in particular the case where the spacetime has no continuous
symmetries and hence a fully 3-spatial-dimensional numerical grid must
be used.  For convenience of exposition I primarily consider the case
where there is precisely one (excised) black hole in each slice; I
briefly discuss possible extensions to multiple black holes in the
conclusions.

Polar spherical coordinates and grids centered on a black hole provide
a natural way to do excision, with an $r = \constant$ excision surface
being a (smooth) grid plane.  However, problems with $z$~axis coordinate
singularities (and the difficulty of generalizing to multiple black holes)
have led most researchers to switch to Cartesian grids over the past
decade.

Unfortunately, it's difficult to do excision on a Cartesian grid:
\begin{enumerate}
\item	\label{excision-case-cube}
	If the excision boundary is a cube, there are severe
	causality problems
(\citet{Scheel2000:Santa-Barbara-talk,
Calabrese2003:excision-and-summation-by-parts,
Lehner-2003:Kerr-cubical-excision-problems,
Lehner-etal-2004:Kerr-cubical-excision-problems}).
	For Schwarzschild spacetime in Eddington-Finkelstein
	(spacelike) coordinates, the excised region must be
	{\em very\/} small to keep the evolution causal.  For
	Kerr spacetime in Kerr-Schild coordinates this problem
	is even worse, and for most spin parameters there is
	{\em no\/} cubical excision region which keeps the
	evolution causal.  For a dynamic spacetime this problem
	is even more severe.
\item	If the excision boundary is some other regular polyhedron
	which better approximates a sphere (such as the 14-sided
	shape obtained by cutting a pyramid off each of a cube's
	corners), the continuum causality problem is less severe,
	but the additional edges and corners considerably
	complicate the construction of a stable finite differencing
	scheme.
\item	If the excision boundary is an irregular \defn{LEGO sphere},
	then at the continuum level (approximating the boundary as a
	sphere centered on the origin), any non-extremal Kerr spacetime
	admits a causal evolution (\citet[figure~3.8]{Thornburg93}).
\footnote{
	 For the Kerr slicing of Kerr spacetime, an
	 $r = \constant$ excision boundary gives a
	 causal evolution whenever the boundary lies
	 between the inner and outer horizons.  This
	 gives a relatively broad range of allowable
	 boundary positions for small spin parameters,
	 narrowing as the spin increases.
	 }
$^,$
\footnote{
	 Since the causality depends only
	 {\em algebraically\/} on the 4-metric,
	 the same property also holds for all
	 ``nearby'' spacetimes in some functional
	 neighborhood of Kerr spacetime.
	 }
{}	However, at the discrete level, the irregularity of the
	boundary makes it very difficult to construct stable finite
	differencing schemes.  (Notably, it's not clear whether the
	powerful numerical schemes of \citet{Calabrese2003a,Calabrese:2003vx}
	can be extended to this case.)  The causality problem of
	case~\ref{excision-case-cube} may also still apply to the
	individual \defn{LEGO sphere} grid points on the excision
	boundary.
\end{enumerate}

In this paper I discuss an alternative technique for doing excision,
which is similar to the use of a polar spherical grid, but with the
$(\theta,\phi)$~angular grid replaced by {\em multiple angular grid patches\/}
which collectively cover $S^2$ in a non-singular manner.  The
corresponding $\{r \times (\text{multiple angular patches}) \}$
coordinate system can be constructed to have no coordinate singularities,
and the multiple-patch grid still has $r = \constant$ as a (smooth)
excision surface.

Like a polar spherical grid, such a multiple-patch grid provides a
smooth $r = \constant$ outer boundary for the imposition of outgoing
radiation conditions, and makes it easy to place the $r = \constant$
shells of grid points nonuniformly in radius so as to give high
resolution close to the black hole while still having the outer
boundary relatively far away.
\footnote{
	 Berger-Oliger mesh refinement
	 (\citet{Berger-1982,Berger84,Berger86,Berger89})
	 provides a much more general and efficient
	 solution here, but it's quite difficult to
	 implement for the Einstein equations (see,
	 for example,
\citet{Choptuik86,Choptuik89,
Liebling-2002:nonlinear-sigma-criticality-via-3D-AMR,
Schnetter-etal-03b,
Bruegmann:2003aw,
Imbiriba-etal-2004:puncture-evolution-FMR},
	 and references therein).
	 }
{}  

Multiple-patch finite difference schemes with smooth excision
surfaces have long been used in computational fluid dynamics
(see, for example, the discussions of
\citet{Rubbert-Lee-in-Thompson-1982:numerical-grid-generation},
\citet[section~II.4]{Thompson-Warsi-Mastin-1985:numerical-grid-generation},
\citet{Chesshire-Henshaw-1990:overlapping-grids-PDEs,
Chesshire-Henshaw-1994:overlapping-grids-conservative-interpolation},
\citet[section~13.4]{Gustafsson95}, \citet{Brown-etal-1997:Overture},
and \citet{Petersson-1999:overlapping-grid-generation}).
However, such schemes have seen relatively little use in numerical
relativity, and that mainly for elliptic problems, either
initial data construction (\citet{Thornburg85,Thornburg87})
or apparent-horizon finding (\citet{Thornburg2003:AH-finding}).
\footnote{
	 Multiple-patch spectral methods have been used
	 in numerical relativity by a number of researchers
	 (see, for example,
\citet{Grandclement-etal-2000:multi-domain-spectral-method,Grandclement02,
Kidder99a,Pfeiffer:2002wt,
Ansorg:2003br,Ansorg:2004vv},
	 and references therein).
	 }

\citet{Bishop-etal-1996:Cauchy-characteristic-matching} have described
a multiple-patch scheme using a pair of overlapping stereographic-coordinate
patches to cover~$S^2$, with field variables in each patch's ghost zone
interpolated (in 2-D) from the other patch.  Their original application
was Cauchy-characteristic matching for radiation boundary conditions,
but the same technique has since been applied for fully-nonlinear
null-cone evolutions (see, for example, \citet{Bishop97b,Gomez97a}).

\citet{Calabrese2003:excision-and-summation-by-parts,
Calabrese-Neilsen-2004:multipatch-email-with-Thornburg}
have described a multiple-patch scheme for the axisymmetric
evolution of a scalar field on a Schwarzschild background.
They use two overlapping patches: a polar spherical inner patch
(with a smooth $r = \constant$ excision surface) and a cylindrical
outer patch.  Ghost-zone values of all the dynamical fields in each
patch's ghost zone are set by (2-D) bilinear interpolation from the
other patch.  Their discretization (\citet{Calabrese2003a,Calabrese:2003vx})
is based on finite difference operators which preserve discrete energy
norms within each patch, maximally dissipative boundary conditions,
and added artificial dissipation to ensure stability.  Their scheme
gives long-term stable evolutions with 2nd~order overall (global)
accuracy;
\footnote{
\label{footnote-accuracy-of-FD-of-interp-data}
	 Although bilinear interpolation has an
	 $O \bigl( (\Delta x)^2 \bigr)$ truncation error,
	 this error is non-smooth at the interpolation
	 points (\citet[appendix~F]{Thornburg98}),
	 so spatial derivatives of the field variables
	 are generically only 1st~order accurate at
	 grid points where interpolated values enter
	 into the derivative molecules.  However,
	 in practice this doesn't seem to seriously
	 affect the overall (global) convergence of
	 the scheme
	 (\citet{Calabrese-Neilsen-2004:multipatch-email-with-Thornburg}).
	 I discuss this point in more detail in
	 section~\ref{sect-finite-differencing}.
	 }
{} a variant discretization gives 3rd~order overall accuracy.

\citet{Tiglio-2003:Penn-State-talk,Reula-2003:Trieste-talk,
Tiglio-2004:multipatch-email-with-Thornburg,
Lehner-Reula-Tiglio-2004:multipatch-scalar-field-Kerr-background}
have described a different multiple-patch scheme for studying scalar
field tails in Schwarzschild or Kerr spacetime.  They use an
\defn{internal boundaries} scheme, where adjacent patches just touch
(rather than overlapping).  By using a symmetric hyperbolic form of
the PDEs, the dynamical fields can be unambiguously decomposed into
ingoing and outgoing modes crossing each interpatch boundary.  By
adding penalty terms to the field equations, the difference between
the value of each ingoing mode at a patch boundary and its value
as an outgoing mode in the adjacent patch, can be driven to zero.

In this paper I describe a multiple-patch excision scheme for the
full nonlinear vacuum $3+1$ Einstein equations.  In terms of past
multiple-patch schemes in numerical relativity, this scheme most
resembles those of \citet{Gomez97} and
\citet{Calabrese2003:excision-and-summation-by-parts},
in that I couple the patches by ghost-zone interpolation of all the
dynamical fields.  However, my scheme uses patches with may either
overlap or just touch, and there are many other differences in detail.

I first implemented the \defn{$g$-$K$} ADM form of the $3+1$ Einstein
equations (\citet{York79}), but it's now well known that these are
ill-posed even in the absence of boundary conditions (see, for example,
the numerical tests of \citet{Alcubierre2003:mexico-I}), and I saw
rapid error growth near the interpatch boundaries, with evolutions
crashing in $\ltsim 100m$ (\citet{Thornburg2000:multiple-patch-evolution}).
I now use the Baumgarte-Shapiro-Shibata-Nakamura (BSSN) form of the
$3+1$ Einstein equations first introduced by
\citet{Nakamura87,Nakamura89,Shibata95,Nakamura99a,Shibata99a},
and popularized by \citet{Baumgarte99}.  In particular, I use the
``actively forced~$A$'' variant described by \citet{Alcubierre99d}.

In this paper I also present initial numerical results for the
BSSN~system, evolving octant-symmetry Kerr spacetime with high accuracy
for $\gtsim 1000m$, and lasting for $t \gtsim 1500m$ before crashing.
These evolutions are unstable at the outer boundary (where I have
only partially implemented Sommerfeld outgoing-radiation boundary
conditions), but they show no signs of any instabilities caused by
the multiple-patch scheme.  There are also no signs of any instabilities
at the inner (excision) boundary.


\subsection{Notation}

I generally follow the sign and notation conventions of \citet{Wald84}.
I use the Penrose abstract-index notation, with indices $a$--$z$ running
over both Cartesian coordinates $\bar{x}^i \equiv (x,y,z)$ in a spacelike
$3+1$~slice, and the generic patch coordinates $x^i \equiv (r,\rho,\sigma)$
defined in section~\ref{sect-multipatch/choice-of-patches+coords}.

$g_{ij}$ is the 3-metric in the slice, $g$ is its determinant,
and $\del_i$ is the associated covariant derivative operator.
$K_{ij}$ is the extrinsic curvature of the slice
(I use the sign convention of \citet{York79}, not that of \citet{Wald84})
and $K \equiv K_i{}^i$ is its trace.
\defn{$N$-D} abbreviates \defn{$N$-dimensional}, and specifically refers
only to spatial dimensions (i.e. the temporal dimension is never included).

I use {\sf sans-serif} font for the names of patches, ghost zones,
and (in appendix~\ref{app-ghost-zone-sync-details}) ghost zone widths
and patch interpolation regions.
In section~\ref{sect-multipatch/change-of-tensor-basis} I use
indices $abc$ and $ijk$ for the $(r,\rho,\sigma)$ coordinates of a pair
of neighboring patches~$\p$ and~$\q$ respectively, and I use the
notation~$f(\p)$ for the grid function $f$ in the local coordinate basis
of patch~$\p$.

In the context of a particular angular patch boundary, I use
angular coordinates $(\perp,\parallel)$ which are
$(\text{perpendicular},\text{parallel})$ to the boundary.
In appendix~\ref{app-ghost-zone-sync-details} I also use the
corresponding integer grid-point coordinates and grid-function
array indices $(\iperp,\ipar)$, and I use a notation inspired by
the \Cplusplus{} programming language, where $(\p.\iperp,\p.\ipar)$
refers to the $(\iperp,\ipar)$~coordinates of patch~$\p$.

In appendix~\ref{app-outer-BCs} I use overbars (as in $\bar{\cphi}$)
to refer to field variables in (strictly speaking, their coordinate
components with respect to) the Cartesian coordinates.


\section{Design of the Multiple-Patch System}
\label{sect-multipatch}

In this section I discuss a variety of design issues which arise
when trying to design and implement such an
$\{ r \times (\text{multiple angular patches}) \}$ numerical scheme
for the $3+1$~vacuum Einstein equations.


\subsection{Treatment of Interpatch Boundaries}
\label{sect-multipatch/interpatch-boundaries}

A fundamental design question for any multiple-patch scheme is how
the interpatch boundaries should be treated in the numerical scheme,
or equivalently, how the patches should be coupled together.

In terms of finite differencing, the usual ghost-zone scheme is
conceptually simple: each patch's nominal grid is surrounded by a
molecule-radius \defn{ghost zone} of extra grid points on each side
(face) of the patch.  Values of all the dynamical fields in the
ghost zones are computed (\defn{synchronized}) by interpolation
from the adjacent patches.  Finite differencing can then be done
at all the nominal grid points without further concern for the
interpatch boundaries.  This scheme can be used with essentially
any formulation of the Einstein equations, and also generalizes to
elliptic PDEs in a reasonably straightforward and efficient manner
(\citet{Thornburg85,Thornburg87,Thornburg2003:AH-finding}).
The main problem with the ghost-zone scheme is that there's
relatively little analytical theory regarding its stability,
although \citet{Olsson-Petersson-1996:overlapping-grid-stability}
have obtained some promising results for 1-D model problems,
and \citet{Starius-1980:composite-mesh-FD-for-hyperbolic-PDE}
has proved stability for a 1-D model problem (as well as performed
numerical experiments using the 2-D shallow-water equations).

In contrast, the \defn{internal-boundaries} scheme used by
\citet{Tiglio-2003:Penn-State-talk,Reula-2003:Trieste-talk,
Tiglio-2004:multipatch-email-with-Thornburg,
Lehner-Reula-Tiglio-2004:multipatch-scalar-field-Kerr-background}
requires a formulation of the Einstein equations in which the modes
crossing each internal boundary can be cleanly separated.  This
must be done everywhere on each interpatch boundary, including
(in general) in strong-field regions.  Although this scheme does
generalize to elliptic PDEs, this generalization is quite inefficient
in multiple spatial dimensions.
\footnote{
	 Briefly, in this scheme nonlinear elliptic PDEs
	 are first linearized in the Newton-Kantorovich sense
	 (\citet[appendix~C]{Boyd00}).  Linear elliptic PDEs
	 are then solved by introducing homogeneous solutions
	 where all the interpatch field values vanish,
	 and one particular solution for each interpatch
	 boundary point.  Thus in a single spatial dimension,
	 the number of particular solutions is just twice
	 the number of patches.  But in multiple spatial
	 dimensions, it's twice the number of
	 interpatch-boundary {\em grid points\/},
	 which is generally quite large.
	 }
{}  The main advantage of the internal-boundaries scheme is its
amenability to mathematical analysis at both the continuum and the
finite differencing levels (see, for example,
\citet{Carpenter-etal-1999:high-order-multiple-patch-FD}).
Also, potentially-destabilizing
feedback loops from one patch to another and back again only occur
due to nonlinear effects coupling the oppositely-directed modes,
whereas in the ghost-zone scheme all the field variables participate
directly in such feedback loops.

Because of its simplicity, and for historical reasons, I have chosen
the ghost-zone scheme for this work.


\subsection{Choice of Angular Patches and Coordinates}
\label{sect-multipatch/choice-of-patches+coords}

There are a number of ways to cover $S^2$ with nonsingular coordinate
patches, of which two seem particularly interesting:

One possibility (\citet{Bishop-etal-1996:Cauchy-characteristic-matching})
is to use a pair of stereographic-coordinate patches, one covering the
northern hemisphere and one the southern.  This has the advantage of
relative simplicity (there is only a single interpatch boundary), and
the eth formalism (\citet{Gomez97}) provides an elegant way to represent
and manipulate tensor fields on $S^2$.

However, these coordinates have relatively large coordinate distortion
near the equator, which makes finite differencing somewhat less accurate
there.  Another potential problem is the coordinate choice on each patch:
If Cartesian stereographic coordinates are used on each patch, then the
grids overlap irregularly near the equator (requiring 2-D interpatch
interpolation there), while if polar stereographic coordinates are
used on each patch, then there are coordinate singularities at the
north and south poles.

Another possibility is to use 6~patches, covering neighborhoods of
the $\pm x$, $\pm y$, and $\pm z$~axes respectively.  This has the
disadvantage of relative complexity: there are many interpatch
boundaries, and there are also corners where 3~patches meet.
However, this system has the advantage of relatively low coordinate
distortion, yielding accurate finite differencing.  Also, if the
coordinates are suitably chosen (I describe this in detail below),
it's possible to have adjacent patches always share a common angular
coordinate, so only 1-D interpatch interpolation is needed.

Because of the lower coordinate distortion, and the simplicity of
only needing 1-D interpatch interpolation, I use a 6-patch system
here.  In more detail, I use the \defn{inflated-cube} angular
6-patch system described in~\citet{Thornburg2003:AH-finding}:
Given Cartesian coordinates $(x,y,z)$, with the excised (black hole)
region at the origin, I define 3~angular coordinates on $S^2$ based
on rotation angles about the $xyz$~coordinate axes:
\begin{equation}
							\label{eqn-mu-nu-phi}
\renewcommand{\arraystretch}{1.25}
\begin{array}{c@{}l@{}l}
\mu	& {} \equiv \text{rotation angle about the $x$ axis}
	& {} = \arctan(y/z)						\\
\nu	& {} \equiv \text{rotation angle about the $y$ axis}
	& {} = \arctan(x/z)						\\
\aphi	& {} \equiv \text{rotation angle about the $z$ axis}
	& {} = \arctan(y/x)						
\end{array}
\end{equation}
where all the arctangents are 4-quadrant based on the signs of $x$, $y$,
and $z$.  I then define 6~coordinate patches covering neighborhoods
of the $\pm z$, $\pm x$, and $\pm y$ axes, using the generic (angular)
patch coordinates $(\rho,\sigma)$ defined by
\begin{equation}
\renewcommand{\arraystretch}{1.25}
\begin{tabular}{l@{}l}
$\pm z$	& ~patch has $(\rho,\sigma) = (\mu,\nu)$			\\
$\pm x$	& ~patch has $(\rho,\sigma) = (\nu,\aphi)$			\\
$\pm y$	& ~patch has $(\rho,\sigma) = (\mu,\aphi)$			
\end{tabular}
						\label{eqn-patch-rho-sigma}
\end{equation}

Notice that each patch's $(\rho,\sigma)$ coordinates are nonsingular
throughout a neighborhood of the patch, and that adjacent patches
always share the (common) angular coordinate perpendicular to their
mutual boundary.  The name \defn{inflated-cube} comes from another
way to visualize these patches and coordinates:  Imagine an $xyz$~cube
with $xyz$~grid lines painted on its face.  Now imagine the cube to be
flexible, and inflate it like a balloon, so it becomes spherical in
shape.  The resulting coordinate lines will closely resemble those
for $(\mu,\nu,\aphi)$ coordinates.

\begin{figure}[tbp]
\begin{center}
\includegraphics[width=100mm,trim=45mm 60mm 20mm 40mm]{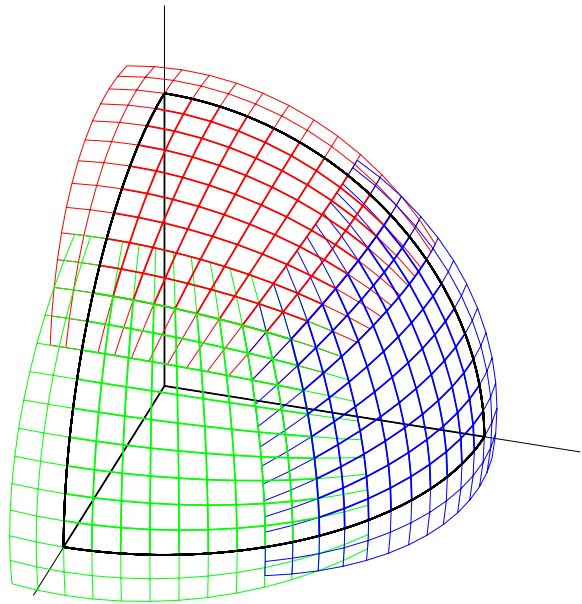}
\end{center}
\caption[Illustration of Multiple Grid Patches]
	{
	This figure shows a multiple-grid-patch system
	covering the $(+,+,+)$~octant of $S^2$ with 3~patches,
	at an angular resolution of $\Delta\rho\sigma = 5^\circ$.
	The $+z$, $+x$, and $+y$ patches
	are shown in red, green, and blue respectively.
	The patch's nominal grids (shown in thick lines)
	just touch; their ghost zones (which are 2~points wide,
	and are shown in thin lines) overlap.
	}
\label{fig-3-patch}
\end{figure}

This set of 6~patches covers $S^2$ without coordinate singularities.
Alternatively, if the spacetime has $z \leftrightarrow -z$ reflection
symmetry about the coordinate origin, then the 5~patches $+z$, $\pm x$,
and $\pm y$ cover the $+z$~hemisphere of $S^2$.  Similarly, suitable
sets of~4 or 3~patches may be used to cover quadrants or octants.
Figure~\ref{fig-3-patch} shows an example of a 3-patch system
covering the $(+,+,+)$~octant of $S^2$.

Defining the usual radial coordinate $r \equiv (x^2 + y^2 + z^2)^{1/2}$,
the corresponding 6-patch (or 5-, 4-, or 3-patch) $(r,\rho,\sigma)$
coordinates cover all of $\Re^3$ (or the corresponding subsets) with
no singularities except at the origin (which is excised).


\subsection{Choice of Tensor Basis}
\label{sect-multipatch/choice-of-tensor-basis}

For numerical solution I decompose the $3+1$~Einstein equations into
coordinate components with respect to a tensor basis.  This raises
the question of what basis should be used in each patch.  There are
two
natural approaches:
\begin{description}
\item[Use the same (typically Cartesian) basis in each patch]
\jtitembreak
	This approach avoids the need for any change of basis in the
	interpatch interpolation.  By using a Cartesian basis, it may
	also make it easier to interface with other numerical relativity
	software which uses this type of tensor basis.

	However, this approach has the disadvantage that a nontrivial
	(frame) transformation is required at each grid point to
	convert grid finite differences into approximate coordinate
	partial derivatives.  This may both slow down the code, and
	possibly introduce time evolution instabilities, or at least
	complicate stability analysis.  It also makes it difficult
	to reuse existing software modules for the evolution.

	Using a Cartesian basis also makes it harder to treat the
	radial coordinate specially for such things as special
	finite differencing near the excision boundary, outer
	boundary conditions, etc.
\item[\jtbold
      Use each patch's own $(r,\rho,\sigma)$ coordinates
      to define its tensor basis]
\jtitembreak
	Because this approach uses a {\em different\/} tensor basis
	in each patch, it requires a change of basis in the interpatch
	interpolation.  (This is discussed in detail in
	section~\ref{sect-multipatch/change-of-tensor-basis}.)
	However, this extra computation is only needed at the
	patch boundaries -- in the interior of each patch there
	is no extra overhead, and existing unigrid software can
	potentially be reused with little or no change.

	This approach also makes it easy to treat the radial
	coordinate specially in finite differencing and/or in
	the continuum formulation of the equations.
\end{description}

In this work I use the second approach: in each patch I use that patch's
own $(r,\rho,\sigma)$ coordinates to define the tensor basis.

\subsection{Synchronizing Ghost Zones}
\label{sect-multipatch/synchronizing}

As described in section~\ref{sect-multipatch/interpatch-boundaries},
I use the usual \defn{ghost zone} technique for handling finite
differencing near the patch boundaries.  I refer to the non--ghost-zone
part of a patch's grid as its \defn{nominal} grid, and to the process
of computing values for a set of grid functions in all ghost zones
of all patches as \defn{synchronizing} these grid functions.  This
must be done for all grid functions to which finite difference
molecules will be applied, \ie{} in practice, for all grid functions
representing dynamical field variables.

Each patch is a rectangular solid (``cuboid'') in its own $(r,\rho,\sigma)$
coordinates, so it has 6~ghost zones: 2~radial (inner and outer)
and 4~angular ($\rho_{\min}$, $\rho_{\max}$, $\sigma_{\min}$, and
$\sigma_{\max}$).  There are three types of ghost zones, each with
corresponding synchronization techniques:
\begin{description}
\item[A radial ghost zone]
	is synchronized by extrapolation from the nominal grid.  I
	describe this in detail in appendix~\ref{app-ghost-zone-sync-details}.
\item[A \defn{symmetry} (angular) ghost zone]
	is one where the patch boundary is a discrete symmetry plane
	of the spacetime, such as the $x=0$, $y=0$, or $z=0$ plane
	in the octant-symmetry example of figure~\ref{fig-3-patch}.
	In this case the symmetry operation defines a mapping from
	the ghost zone into the nominal grid of some (possibly other) patch,
	so the ghost-zone grid function values can be computed by
	copying from the symmetry-image grid points.
\item[An \defn{interpatch} (angular) ghost zone]
	is one where the ghost zone lies within the nominal grid
	of some other patch.
\footnote{
	 To ensure that this is always so, adjacent patches'
	 nominal grids must either overlap, just touch,
	 or be separated by a gap of at most one grid spacing.
	 }
{}	In this case the ghost-zone grid function values can be
	computed by interpolating from the neighboring patch.
	Because all patches share a common radial coordinate,
	and (by construction) adjacent patches share the angular
	coordinate perpendicular to their mutual boundary, the
	(1-D) interpatch interpolation is done (independently
	in each line of constant perpendicular coordinate and $r$)
	only in the direction parallel to the boundary.
\end{description}

Because off-centered interpolations are relatively inaccurate,
and have large phase errors which may lead to finite differencing
instabilities in the time evolution, I try to keep the interpatch
interpolations as centered as possible.

A major complication in ghost zone synchronization is that ghost
zones have {\em corners\/}, \ie{} there are ghost-zone points which
are outside the nominal grid in more than one dimension.  It's quite
tricky to ensure that the corner ghost-zone points are properly updated
while still keeping the interpatch interpolations as centered as
possible near the patch corners.  I use the 3-phase algorithm of
\citet[appendix~A]{Thornburg2003:AH-finding}, generalized to also
handle radial ghost zones and non-scalar grid functions, to synchronize
ghost zones.  I describe the resulting algorithm in detail in
appendix~\ref{app-ghost-zone-sync-details}.  It turns out that if
the interpatch-interpolation molecule is 4-point or smaller, then
all the interpatch interpolations can in fact be kept centered.
If the interpatch-interpolation molecule is 5-point or larger,
then all the interpatch interpolations can still be kept centered
except in the immediate neighborhood of a radial line where 3~patches
meet.


\subsection{Change of Tensor Basis}
\label{sect-multipatch/change-of-tensor-basis}

In numerical relativity we need to deal with tensors and other non-scalar
grid functions, so interpatch symmetry operations and interpolations
entail a change of basis (step~\ref{step-change-of-basis} in the
synchronization algorithm of figure~\ref{fig-ghost-zone-sync-algorithm}).

Without loss of generality I assume that all the fields are known at a
point (event) in some patch $\p$ in some other patch~$\q$'s $(r,\rho,\sigma)$
basis~$\{ x^i(\q) \}$, and we wish to transform them to patch~$\p$'s
$(r,\rho,\sigma)$ basis~$\{ x^a(\p) \}$.  I define the transformation
matrices
\begin{subequations}
\begin{eqnarray}
X^a{}_i
	& = &	\ddfrac{\partial x(\p)^a}{\partial x(\q)^i}		\\
Y^i{}_a
	& = &	\ddfrac{\partial x(\q)^i}{\partial x(\p)^a}		\\
Y^i{}_{ab}
	& = &	\ddfrac{\partial^2 x(\q)^i}{\partial x(\p)^a \partial x(\p)^b}
\end{eqnarray}
\end{subequations}
These are straightforward to compute analytically from the coordinate
definitions~\eqref{eqn-mu-nu-phi}.

Of the BSSN field variables, $\alpha$ and $K$ are scalars, $\beta^i$
is a tensor and transforms as
\begin{equation}
\beta^a(\p) = X^a{}_k \beta^k(\q)
\end{equation}
and $\cphi$ is (proportional to) the logarithm of a tensor density
and transforms as
\begin{equation}
\cphi(\p) = \cphi(\q) + \tfrac{1}{6} \log \left|Y^\parallel{}_\parallel\right|
							\label{eqn-cphi-cxform}
\end{equation}
where $\parallel$ refers to each patch's angular coordinate parallel
to $\p$ and $\q$'s mutual interpatch boundary.
$\tilde{g}_{ij}$ and $\tilde{A}_{ij}$ are tensor densities, and
transform as
\begin{subequations}
\begin{eqnarray}
\tilde{g}_{ab}(\p)
	& = &	\left|Y^\parallel{}_\parallel\right|^{-2/3}
		Y^i{}_a Y^j{}_b \tilde{g}_{ij}(\q)			\\
\tilde{A}_{ab}(\p)
	& = &	\left|Y^\parallel{}_\parallel\right|^{-2/3}
		Y^i{}_a Y^j{}_b \tilde{A}_{ij}(\q)			
\end{eqnarray}
\end{subequations}

$\tilde{\Gamma}^i$ isn't a tensor or tensor density, so it transforms
in a more complicated manner (I outline the derivation of this in
appendix~\ref{app-Gamma-tilde-u-cxform}),
\begin{eqnarray}
\tilde{\Gamma}^a(\p)
	& = &	\left|Y^\parallel{}_\parallel\right|^{2/3}
		X^a{}_k \tilde{\Gamma}^k(\q)
		+ X^a{}_k Y^k{}_{bc} \tilde{g}^{bc}(\p)
							\nonumber	\\
	&   &	\quad
		{}
		- 2 \left|Y^\parallel{}_\parallel\right|^{2/3}
		    X^a{}_k \tilde{g}^{k\ell}(\q) \partial_\ell \cphi(\q)
		+ 2 \tilde{g}^{ab}(\p) \partial_b \cphi(\p)
						\label{eqn-Gamma-tilde-u-cxform}
\end{eqnarray}
The first two terms here are straightforward to implement, but the
3rd~and 4th~terms are problematic in several ways:
\begin{itemize}
\item	The 3rd~term of~\eqref{eqn-Gamma-tilde-u-cxform} involves
	the inverse conformal metric $\tilde{g}^{k\ell} (\q)$.
	Since this is a patch-$\q$ quantity, it's presumably only
	known on the patch-$\q$ grid, and so must be interpolated
	to the (incomeasurate) patch-$\p$ grid.  This requires either
	keeping $\tilde{g}^{ij}$ as a grid function so that it can
	be interpolated (instead of the more economical-of-memory
	alternative of only storing its value at the current grid point),
	or interpolating $\tilde{g}_{ij}$ and then computing
	$\tilde{g}^{ij}$ from that.  In my code I do the latter.
\item	The 3rd~term of~\eqref{eqn-Gamma-tilde-u-cxform} also involves
	$\partial_\ell \cphi(\q)$.  Since $\cphi$ is a dynamic field
	variable, this partial derivative must be computed by finite
	differencing.  This is difficult because this is a patch-$\q$
	derivative, but it's needed at a patch-$\p$ grid point.  The
	computation can be done by either using a differentiating
	interpolator
\footnote{
	 An interpolator generally works by (conceptually)
	 locally fitting a fitting function (usually a
	 low-degree polynomial) to the data points in a
	 neighborhood of the interpolation point, then
	 evaluating the fitting function at the interpolation
	 point.  By evaluating the {\em derivative\/} of the
	 fitting function, the $\partial_\ell \cphi(\q)$ values
	 can be interpolated very cheaply, using only the
	 input $\cphi(\q)$ values which are used anyway for
	 interpolating $\cphi(\q)$ to the patch-$\p$ grid.
	 }
{},	or (less efficiently) by storing the $\partial_\ell \cphi(\q)$
	as grid functions and interpolating them directly.
	(This computation is being done at a point which lies
	within patch~$\q$'s nominal grid, so there's no problem
	in computing $\partial_\ell \cphi(\q)$.)  For historical
	reasons, I use the latter technique in my current code.
\item	The 4th~term of~\eqref{eqn-Gamma-tilde-u-cxform} involves
	$\partial_b \cphi(\p)$, which is particularly difficult to
	compute, because (as in the 3rd~term) this derivative must
	be computed by finite differencing, and this must be done
	at a point which lies in patch~$\p$'s ghost zone.
	There are two ways to compute this term:
	\begin{itemize}
	\item	$\partial_b \cphi(\p)$ can be computed by
		straightforward finite differencing of $\cphi(\p)$
		if $\cphi$ has a ghost zone twice as wide as that of
		$\tilde{\Gamma}^i$ and the other BSSN field variables,
		so that the computation point (in $\tilde{\Gamma}^i$'s
		ghost zone) is still guaranteed to be surrounded by
		at least a molecule-sized neighborhood of $\cphi(\p)$
		grid points.
	\item	Alternatively, using the analytically-known
		$\cphi(\p) \leftrightarrow \cphi(\q)$
		transformation~\eqref{eqn-cphi-cxform},
		$\partial_b \cphi(\p)$ can be rewritten
		in terms of the $\partial_\ell \cphi(\q)$,
		which can be computed using the methods
		described above for the 3rd~term
		of~\eqref{eqn-Gamma-tilde-u-cxform}.
\footnote{
	 Actually this is only needed for the perpendicular
	 derivative $\partial_\perp \cphi(\p)$, since the
	 radial and parallel derivatives $\partial_r \cphi(\p)$
	 and $\partial_{\parallel\!(\p)} \cphi(\p)$ are taken in
	 directions parallel to the ghost zone, and so can be
	 computed by standard finite differencing techniques
	 without using any $\cphi(\p)$ data from outside the
	 ghost zone.
	 }
{}		This avoids having to introduce double-width
		ghost zones for~$\cphi$.
	\end{itemize}
	Because it minimizes the amount of interpolation to be done,
	and seems to give a slightly simpler and more easily understood
	numerical scheme, I use the double-width ghost zone approach
	for all the numerical results presented here.  In particular
	(given the 5-point finite difference molecules described in
	section~\ref{sect-finite-differencing}), I use angular ghost
	zones which are 2~grid points wide for most grid functions,
	and 4~grid points wide for~$\cphi$.

	Although this approach is conceptually simple, it turns out
	to be somewhat awkward to retrofit to an existing code, as
	large parts of the code generally implicitly assume that the
	ghost-zone width is an inherent property of a grid, rather
	than (potentially) varying from one grid function to another.
\end{itemize}


\section{Finite Differencing}
\label{sect-finite-differencing}

In this section I describe the finite differencing scheme used to
obtain the numerical results presented in section~\ref{sect-results}.
This is a straightforward generalization to 3-D of the 1-D finite
differencing scheme I described in~\citet{Thornburg99}.

To allow higher resolution near the black hole while allowing the
outer boundary to be placed relatively far away, I place the $r = \constant$
shells of grid points non-uniformly in $r$: they are uniformly spaced
in a new (dimensionless) radial coordinate $\wr = \wr(r)$.  The
implementation of this nonuniform gridding is identical to that
of~\citet{Thornburg99}, with the parameters $r_0 = 1.5m$, $a = \infty$,
$b = 5m$, and $c = 100m$.  For these parameters $\wr$ qualitatively
resembles a logarithmic radial coordinate in the inner part of the
grid, and a uniform radial coordinate in the outer part of the grid.

I use the method of lines, with a low-storage variant of the
classical 4th~order Runge-Kutta time integrator
(\citet{Blum-1962:low-storage-Runge-Kutta,
Williamson-1980:low-storage-Runge-Kutta}).
For my nonuniform-grid parameters and coordinate conditions
(section~\ref{sect-results/coordinate-conditions}),
the empirical CFL limit of my code is a Courant number of
$\Delta t/\Delta w = 0.63 \pm 0.01$; I use $\Delta t/\Delta w = 0.5$
for all the numerical results presented here.

For the spatial finite differencing, I first transform all
$(r,\rho,\sigma)$~coordinate partial derivatives into
$(\wr,\rho,\sigma)$~coordinate partial derivatives (for example,
$\partial_r f = (\partial \wr / \partial r) \partial_\wr f$)
for any field variable $f$).  I then approximate the
$(\wr,\rho,\sigma)$~derivatives by the usual centered 5-point
4th~order 1st~or 2nd~derivative molecules as appropriate, except
that for the shift vector advection derivatives I use off-centered
5-point molecules, upwinded by 1~grid point in the radial direction
based on the sign of $\beta^r$ (which is always positive for the
numerical results presented here).

At the inner and outer grid boundaries I use 4th~order Lagrange
polynomial extrapolation of the nominal-grid field variables to fill
in values in the radial ghost zones (steps~\ref{step-radial-extrap-1}
and~\ref{step-radial-extrap-2} in the ghost-zone synchronization
algorithm of figure~\ref{fig-ghost-zone-sync-algorithm}).
Because of the upwind derivatives in the radial direction, the
ghost zones are 2(3)~grid points wide at the inner(outer) boundary
for most grid functions, or 4(6)~grid points wide for~$\cphi$.

I use 5th~order (6-point) Lagrange polynomial interpolation for
all the interpatch interpolations (step~\ref{step-interpatch-interp}
in the ghost-zone synchronization algorithm of
figure~\ref{fig-ghost-zone-sync-algorithm}).  This gives an
$O \bigl( (\Delta x)^6 \bigr)$ truncation error for all the
interpolated field variables.  However, these errors -- and thus
the interpolated field variables themselves -- are {\em non-smooth\/}
at the interpolation points (\citet[appendix~F]{Thornburg98}).
Because of this non-smoothness, the consistency argument of
\citet{Choptuik91} does {\em not\/} apply here: taking numerical
derivatives of the interpolated field variables {\em does\/} locally
lower the order of accuracy.

For the ADM equations, taking numerical 2nd~derivatives of the
interpolated field variables gives 4th~order accuracy, and my
numerical tests (omitted here in the interests of brevity)
confirm that the scheme attains this.  However, for the BSSN
equations the $\tilde{\Gamma}^i$ interpatch change-of-basis
transformation~\eqref{eqn-Gamma-tilde-u-cxform} involves numerical
1st~derivatives of the BSSN conformal factor~$\cphi$, so the
results of this transformation have non-smooth 5th~order errors
(as well as the usual smooth 4th~order errors).  Taking numerical
2nd~derivatives of $\tilde{\Gamma}^i$ in the BSSN evolution and
constraint equations thus lowers the local accuracy of my scheme
to 3rd~order near interpatch boundaries.
\footnote{
	 This could be raised to 4th~order by using a
	 higher-order or smoother (Hermite or spline)
	 interpatch interpolation, but I have only made
	 limited trials of this thus far.
	 }

The numerical results presented in section~\ref{sect-results/convergence}
show that despite the lower local order of accuracy near the interpatch
boundaries, in practice my scheme remains globally 4th~order accurate
everywhere away from the grid and patch boundaries.  Like the analogous
result of \citet{Calabrese2003:excision-and-summation-by-parts,
Calabrese-Neilsen-2004:multipatch-email-with-Thornburg}
(described in footnote~\ref{footnote-accuracy-of-FD-of-interp-data}),
and those of \citet{Imbiriba-etal-2004:puncture-evolution-FMR},
this is in accordance with theoretical arguments
(\citet[page~571]{Gustafsson95}) that under suitable conditions
boundary conditions can be approximated one order lower in accuracy
than interior equations, without affecting the global order of accuracy
of the scheme.

Because the inner (excision) grid boundary (placed at $r = 1.5m$ for
all the numerical results presented here) is spacelike, no further
boundary conditions are needed there, and I use the usual interior
evolution equations to determine the time evolution of the field
variables.  Because I only use 4th~order radial extrapolation,
not 5th~order, the finite differencing is only 3rd~order accurate
for 2nd~derivatives at the inner and outer grid boundaries.  This
doesn't degrade the overall 4th~order evolution accuracy at the
inner boundary because the grid points there are an ``outflow''
boundary with respect to the evolution causality
(\citet{Gustafsson1971:hyperbolic-BC-FD-convergence,
Gary1975:MOL-outflow-BC,
Gustafsson1975:hyperbolic-BC-FD-convergence,Gustafsson79,
Gustafsson1982:hyperbolic-BC-FD-convergence}).
\footnote{
	 If it were desirable, there would be no particular
	 problem with using 5th~order radial extrapolation
	 to avoid the lower order accuracy at the radial
	 boundaries (\citet{Thornburg99}); I have not done
	 this in the present work only for historical reasons.
	 }

At the outer boundary I use the outer boundary conditions described
in appendix~\ref{app-outer-BCs} to determine the time derivatives
of the BSSN field variables.
\footnote{
	 The errors at the outer boundary are dominated by
	 those from the continuum outer boundary conditions
	 (appendix~\ref{app-outer-BCs}), so the 3rd~order
	 finite differencing there is unimportant.
	 }
{}  Unfortunately, there's no reason to think these boundary conditions
are either well-posed or constraint-preserving, and outer boundary
instabilities seem to be the limiting factor for the evolutions I
present in section~\ref{sect-results}.

\citet{Olsson-Petersson-1996:overlapping-grid-stability} found that
some artificial dissipation was necessary to obtain stable evolutions
with a two-patch scheme for a simple (linear) 1-D model problem.
I haven't included any artificial dissipation in my numerical scheme,
though a small amount of dissipation is present inherently due to
the time integrator and the upwind finite differencing of the
shift vector advection terms.  I don't know whether adding
artificial dissipation would improve my scheme's stability.


\section{Numerical Results}
\label{sect-results}

In this section I present various sample results from a prototype
implementation of the numerical scheme described in this paper.
This code is a standalone uniprocessor code, not using any larger
relativity toolkit such as Cactus (\citet{Goodale02a}).
The multiple-patch infrastructure (roughly 10K~lines of \Cplusplus)
was quite difficult to design and debug, though I suspect a
reimplementation would be somewhat simpler.  I found no problems
in combining the multiple-patch scheme with ``relativity'' code
machine-generated from a higher-level tensor form.


\subsection{Initial Data}

For the numerical results presented here, I use initial data for
the lapse, shift, and BSSN field variables which is a Kerr-coordinate
slice of Kerr spacetime, with spin $J/m^2 = 0.6$.  I use this same
slice as the background slice for my outer boundary conditions
(appendix~\ref{app-outer-BCs}).  In Kerr coordinates the horizon
is at $r = (1 + \sqrt{1 - a^2})m = 1.8m$ for this spin; as mentioned
above, I place the inner (excision) grid boundary at $r = 1.5m$.


\subsection{Coordinate Conditions}
\label{sect-results/coordinate-conditions}

I use a time-independent shift vector for all the numerical results
presented here, with $\beta^i$ set to a suitable (position-dependent)
value on the initial slice and not updated thereafter.

I use generalizations of the Bona-Masso slicings (\citet{Bona94b}).
In particular, I use a slicing condition slightly adapted from the
``$K$-driver'' condition of
\citet{Alcubierre02a,Alcubierre2003:hyperbolic-slicing},
\begin{equation}
\partial_t \alpha
	= - \alpha f(\alpha) (\alpha K - \del_i \beta^i)
\end{equation}
with $f(\alpha) = A \alpha^n$.  As discussed by
\citet[section~III.A]{Alcubierre02a}, these slicings have a gauge
propagation speed of
\begin{equation}
c_\text{lapse}
	= \alpha \sqrt{f(\alpha)}
					\label{eqn-Bona-Masso-lapse-speed}
\end{equation}

I made some evolutions with $A = 2$, $n=-1$ (the ``$1 + \log$''
slicing recommended by \citet{Alcubierre02a}), but these suffered
from severe gauge instabilities.  I use $A = 2$, $n=0$ (a variant
of harmonic slicing) for all the results reported here.


\subsection{Numerical Grid Parameters}

I have evolved this initial data with a number of different finite
differencing grids, as shown in table~\ref{tab-grid-parameters}.
All the results presented here are for data with octant symmetry.

\begingroup
\squeezetable
\begin{table}[tbp]
\begin{center}
\begin{tabular}{llcccccccl}
		&		&		&		&
		& \multicolumn{2}{c}{Outer Boundary}
		& \multicolumn{2}{c}{Angular Grid}
		&							\\
		\cline{8-9}
		&	& \multicolumn{3}{c}{Radial Resolution $\Delta r$}
		& \multicolumn{2}{c}{Position}
		& Resolution
						& Patch
		&							\\
		\cline{3-5}
%
Model		& $\Delta\wr$	& Inner		& $\wr=0.12$	& Outer
		& $\wr_{\max}$	& $r_{\max}$
		& $\Delta\rho\sigma$\P{0}	& Overlap
		& Outcome						\\
\hline 
33k-wrmax4	& 0.03		& 0.143$m$	& 0.204$m$	& 2.7$m$
		& \P{0}4	& 248$m$
		& 4.5$^\degree$\P{0}		& $\pm 0$~(1)
		& crash at $t=1296m$					\\
50k-wrmax4	& 0.02		& 0.095$m$	& 0.136$m$	& 1.8$m$
		& \P{0}4	& 248$m$
		& 3.0$^\degree$\P{0}		& $\pm 0$~(1)
		& crash at $t=1719m$					\\
66k-wrmax4	& 0.015		& 0.071$m$	& 0.102$m$	& 1.3$m$
		& \P{0}4	& 248$m$
		& 2.25$^\degree$		& $\pm 0$~(1)
		& crash at $t=1747m$					\\
\hline 
33k		& 0.03		& 0.143$m$	& 0.204$m$	& 2.2$m$
		& \P{0}2	& \P{0}82$m$
		& 4.5$^\degree$\P{0}		& $\pm 0$~(1)
		& crash at $t=1030m$					\\
50k		& 0.02		& 0.095$m$	& 0.136$m$	& 1.5$m$
		& \P{0}2	& \P{0}82$m$
		& 3.0$^\degree$\P{0}		& $\pm 0$~(1)
		& crash at $t=991m$					\\
66k		& 0.015		& 0.071$m$	& 0.102$m$	& 1.1$m$
		& \P{0}2	& \P{0}82$m$
		& 2.25$^\degree$		& $\pm 0$~(1)
		& crash at $t=897m$					\\
75k		& 0.01333	& 0.063$m$	& 0.091$m$	& 1.0$m$
		& \P{0}2	& \P{0}82$m$
		& 2.0$^\degree$\P{0}		& $\pm \tfrac{1}{2}$~(2)
		& crash at $t=1448m$					\\
\hline 
33k-overlap3	& 0.03		& 0.143$m$	& 0.204$m$	& 2.2$m$
		& \P{0}2	& \P{0}82$m$
		& 4.5$^\degree$\P{0}		& $\pm 3$~(7)
		& crash at $t=302m$					\\
50k-overlap3	& 0.02		& 0.095$m$	& 0.136$m$	& 1.5$m$
		& \P{0}2	& \P{0}82$m$
		& 3.0$^\degree$\P{0}		& $\pm 3$~(7)
		& crash at $t=789m$					\\
66k-overlap3	& 0.015		& 0.071$m$	& 0.102$m$	& 1.1$m$
		& \P{0}2	& \P{0}82$m$
		& 2.25$^\degree$		& $\pm 3$~(7)
		& crash at $t=1307m$					\\
\hline 
33k-9overlap	& 0.03		& 0.143$m$	& 0.204$m$	& 2.2$m$
		& \P{0}2	& \P{0}82$m$
		& 4.5$^\degree$\P{0}		& $\pm 9^\degree$\,(5)
		& crash at $t=492m$					\\
50k-9overlap	& 0.02		& 0.095$m$	& 0.136$m$	& 1.5$m$
		& \P{0}2	& \P{0}82$m$
		& 3.0$^\degree$\P{0}		& $\pm 9^\degree$\,(7)
		& crash at $t=789m$					\\
66k-9overlap	& 0.015		& 0.071$m$	& 0.102$m$	& 1.1$m$
		& \P{0}2	& \P{0}82$m$
		& 2.25$^\degree$		& $\pm 9^\degree$\,(9)
		& crash at $t=1042m$					\\
\hline 
33k-wrmax10	& 0.03		& 0.143$m$	& 0.204$m$	& 2.9$m$
		& 10		& 813$m$
		& 4.5$^\degree$\P{0}		& $\pm 0$~(1)
		& crash at $t=859m$					\\
50k-wrmax10	& 0.02		& 0.095$m$	& 0.136$m$	& 1.9$m$
		& 10		& 813$m$
		& 3.0$^\degree$\P{0}		& $\pm 0$~(1)
		& crash at $t=1511m$					\\
66k-wrmax10	& 0.015		& 0.071$m$	& 0.102$m$	& 1.4$m$
		& 10		& 813$m$
		& 2.25$^\degree$		& $\pm 0$~(1)
		& still running at $t = 1867m$				\\
\hline 
\end{tabular}
\end{center}
\caption[Grid Parameters]
	{
	This table shows the grid parameters and run outcomes
	for the various evolutions.  The radial resolution is given
	first as the radial grid spacing in $\wr$, then as the
	radial grid resolution $\Delta r$ at the inner boundary,
	the $\wr = 0.12$ ($r = 2.19m$) position used for
	the angular convergence plots, and the outer boundary.
	The \defn{patch overlap} is specified in one of two ways:
	$\pm m$~(n) or $\pm x^\degree$\,(n).  $\pm m$ or $\pm x^\degree$
	gives the distance of the patches' nominal grid boundaries
	from $45^\degree$ (either $m$~angular grid spacings, or
	$x$~degrees), while $n$ gives the number of perpendicular
	coordinate values common to a pair of overlapping patches.
	For example, for a model with angular grid resolution
	$\Delta\rho\sigma = 3^\degree$,
	\defn{$\pm 3$~(7)} and \defn{$9^\degree$\,(7)} both mean
	that the patches' nominal grid boundaries are at
	$45^\degree \pm 3 \, \Delta\rho\sigma = 45^\degree \pm 9^\degree$,
	so adjacent patches have 7~grid points in common in their
	perpendicular-to-the-boundary direction.  (Thus the
	50k-overlap3 and 50k-9overlap models are actually identical.)
	}
\label{tab-grid-parameters}
\end{table}
\endgroup


\subsection{Diagnostics}

I use two main diagnostics to assess the code's accuracy:
The first is the energy constraint $C \equiv R - K_{ij} K^{ij} + K^2$.
To determine a scale for nonzero values of this, it's useful to also
consider $C_\abs \equiv |R| + |K_{ij} K^{ij}| + K^2$, and its value
$C_\absexact$ computed analytically for my (Kerr) initial data.
$C_\absexact$ is $O(1)$ near the black hole, but falls off $\sim r^{-4}$
at large~$r$.  The \defn{relative energy constraint} $C/C_\absexact$
is a dimensionless measure of the accuracy with which the code is
approximating a solution of the Einstein equations.

My second diagnostic is the error of the BSSN state vector with respect
to its analytical (Kerr) value for my initial data,
\begin{equation}
\delta S \equiv
	(\delta\cphi)^2
	+ \sum_{i \le j} (\delta\tilde{g}_{ij})^2
	+ (\delta K)^2
	+ \sum_{i \le j} (\delta\tilde{A}_{ij})^2
	+ \sum_i (\delta\tilde{\Gamma}^i)^2
\end{equation}
where for each BSSN field variable
$f \in \{ \cphi,\tilde{g}_{ij}, K, \tilde{A}_{ij}, \tilde{\Gamma}^i \}$,
$\delta f \equiv f - f_\text{Kerr}$.

I consider a run to \defn{crash} if any BSSN field variable exceeds
$10^{10}$ in magnitude at any grid point.  Table~\ref{tab-grid-parameters}
lists the times at which this occurs for each run.


\subsection{Convergence}
\label{sect-results/convergence}

\begin{figure}[tbp]
\begin{center}
\includegraphics[width=190mm,trim=10 30 0 0,clip=true]
   {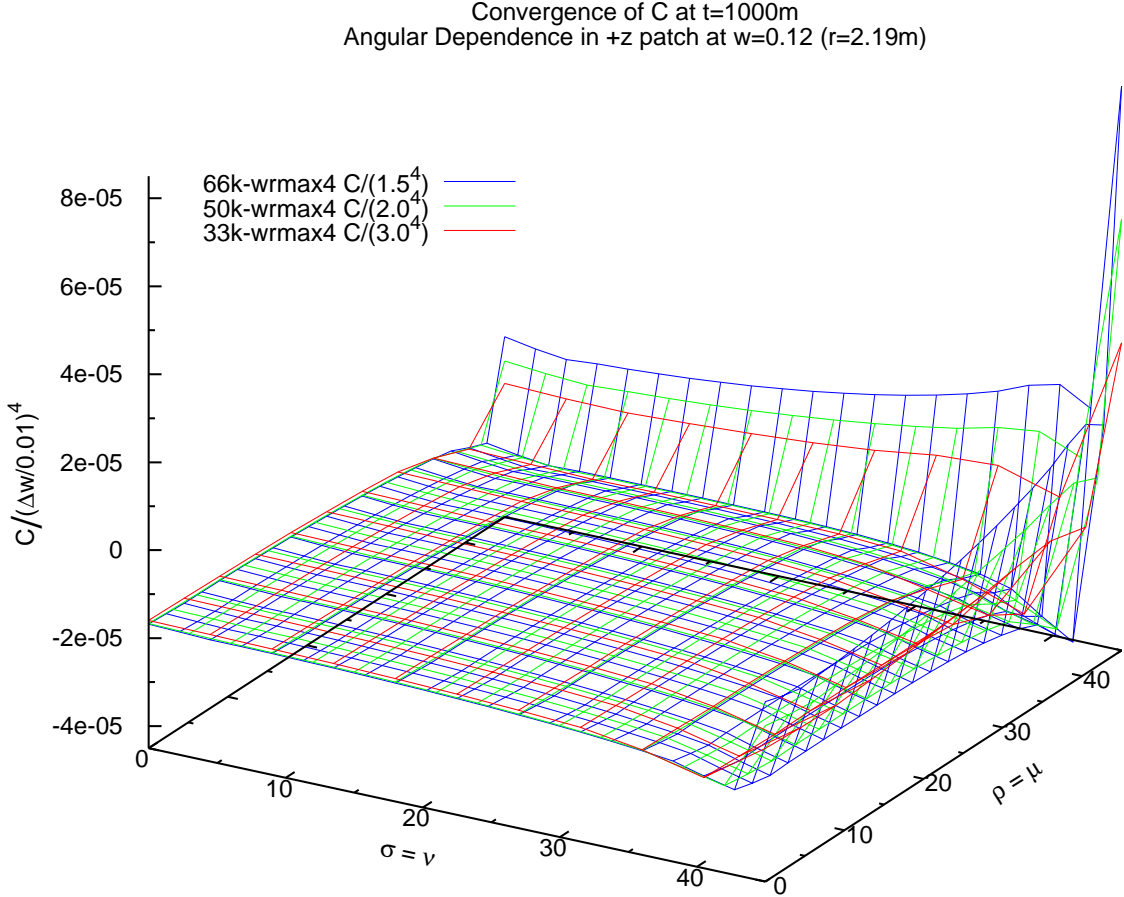}
\end{center}
\caption[Convergence of Energy Constraint at $t=1000m$, Angular Dependence]
	{
	This figure shows the convergence of the
	energy constraint~$C$ for the $+z$ patch of the
	$\wr = 0.12$ ($r = 2.19m$) shell of grid points at $t = 1000m$.
	The horizontal axes show the angular coordinates in degrees
	(so $(0,0)$ is the $z$~axis and $(45,45)$ is the \defn{triple point}
	where the 3~patches meet in figure~\ref{fig-3-patch}).
	The $z$ axis shows the energy constraint (scaled by
	the 4th~power of the resolution) for the 33k-wrmax4,
	50k-wrmax4, and 66k-wrmax4 models.
	}
\label{fig-converge-C-adat-z-t=1000-wrmax=4}
\end{figure}

\begin{figure}[tbp]
\begin{center}
\includegraphics[width=125mm]
   {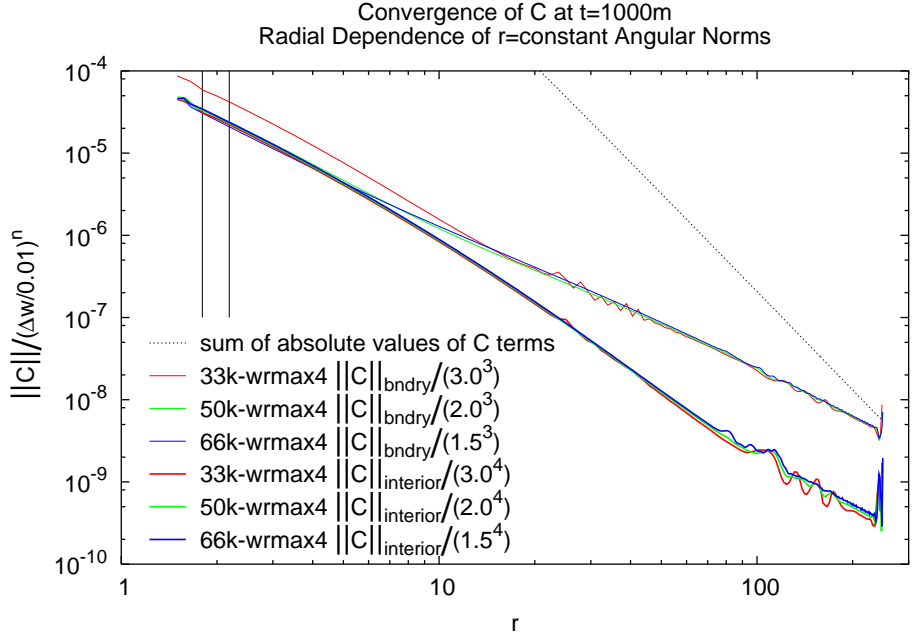}
\end{center}
\caption[Convergence of Energy Constraint at $t=1000m$,
	 Radial Dependence of $r=\constant$ Angular Norms]
	{
	This figure shows the convergence of various $r=\constant$
	angular RMS-norms of the energy constraint~$C$ at $t=1000m$.
	The lower [upper] curves show angular RMS-norms of the
	energy constraint (scaled by the 4th~[3rd]~power of the
	resolution) over the patch-interior [interpatch-boundary]
	grid points in each $r=\constant$ shell; in all cases $\infty$-norms
	are within an order of magnitude of the RMS-norms shown.
	The diagonal dashed line shows $\| C_\absexact \|$.
	The two vertical lines show the horizon position
	and the $\wr = 0.12$ ($r = 2.19m$) position used for
	the angular convergence plots.
	}
\label{fig-converge-C-rdat-t=1000-wrmax=4}
\end{figure}

As discussed in section~\ref{sect-finite-differencing}, in the limit
of infinite resolution and in the absence of outer-boundary effects,
my numerical scheme should give 4th~order convergence for the energy
constraint~$C$ in patch interiors, 3rd~order for $C$ near interpatch
boundaries, and 4th~order convergence everywhere for the state vector
error $\delta S$.  The numerical results are quite close to this for
a considerable period of time.  For example,
figures~\ref{fig-converge-C-adat-z-t=1000-wrmax=4}
and~\ref{fig-converge-C-rdat-t=1000-wrmax=4} show typical convergence
results for the energy constraint~$C$ at $t=1000m$ in the 33k-wrmax4,
50k-wrmax4, and 66k-wrmax4 evolutions.
\footnote{
	 Notice that at large$r$, $C$ falls off considerably
	 more slowly ($\sim r^{-2}$) than $C_\absexact$ (which
	 falls off $\sim r^{-4}$), \ie{} the relative constraint
	 violation $C/C_\absexact$ {\em grows\/} $\sim r^2$.
	 This is due to finite differencing errors involving
	 the $r^2$~factors in the $(r,\rho,\sigma)$ metric
	 components.  It could probably be cured
	 (greatly improving the code's accuracy at large~$r$)
	 by factoring those $r^2$ factors out analytically,
	 but I haven't tried this.
	 }
{}  The best-fitting convergence exponents are about $3.9$~for the
patch interiors and $2.9$~for the interpatch boundaries.

\begin{figure}[tbp]
\begin{center}
\includegraphics[width=190mm,trim=10 25 0 0,clip=true]
   {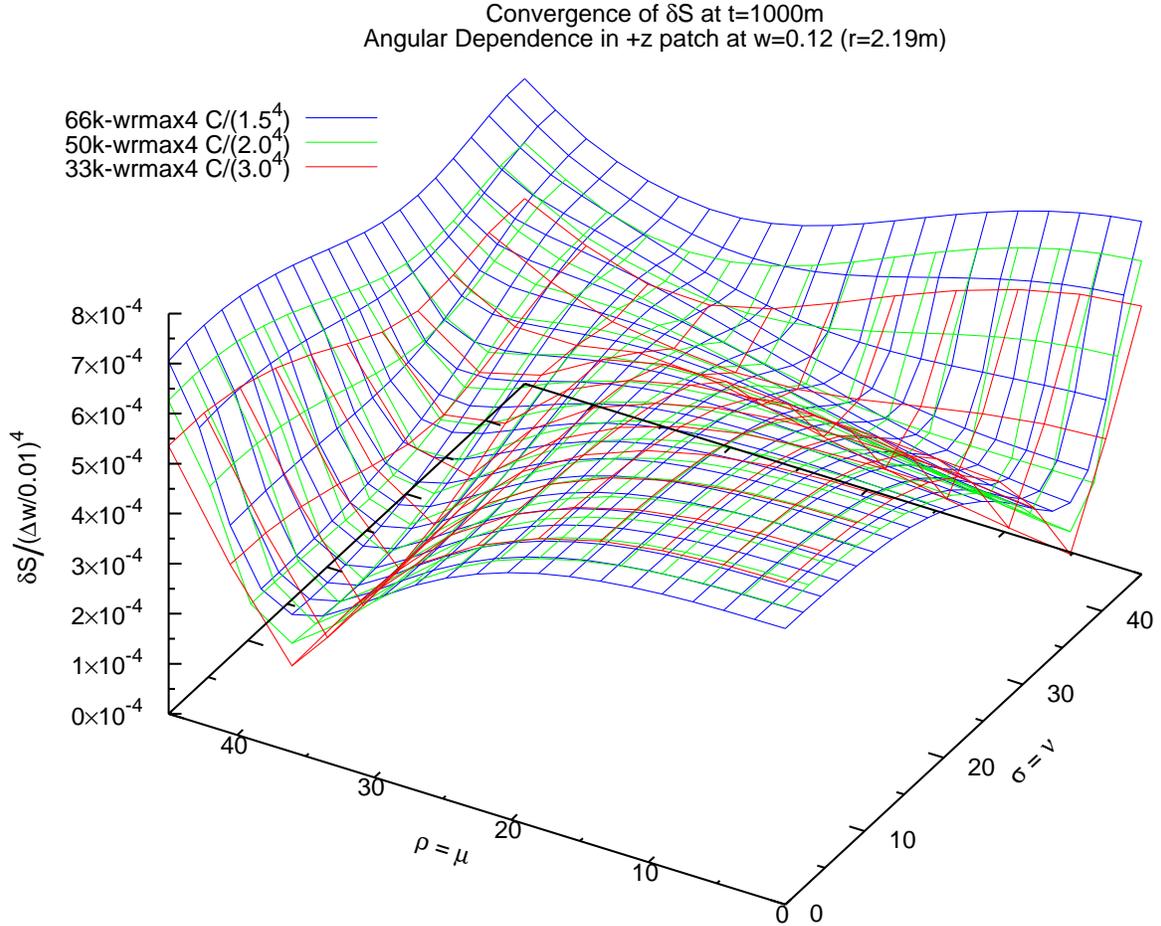}
\end{center}
\caption[Convergence of State Vector Error at $t=1000m$, Angular Dependence]
	{
	This figure shows the convergence of the state vector error~$\delta S$
	for the $+z$ patch of the $\wr = 0.12$ ($r = 2.19m$)
	shell of grid points at $t = 1000m$.
	The horizontal axes show the angular coordinates in degrees
	(so $(0,0)$ is the $z$~axis and $(45,45)$ is the \defn{triple point}
	where the 3~patches meet in figure~\ref{fig-3-patch}).
	The $z$ axis shows the state vector error
	(scaled by the 4th~power of the resolution)
	for the 33k-wrmax4, 50k-wrmax4, and 66k-wrmax4 models.
	}
\label{fig-converge-delta-sv-adat-z-t=1000-wrmax=4}
\end{figure}

\begin{figure}[tbp]
\begin{center}
\includegraphics[width=125mm]
   {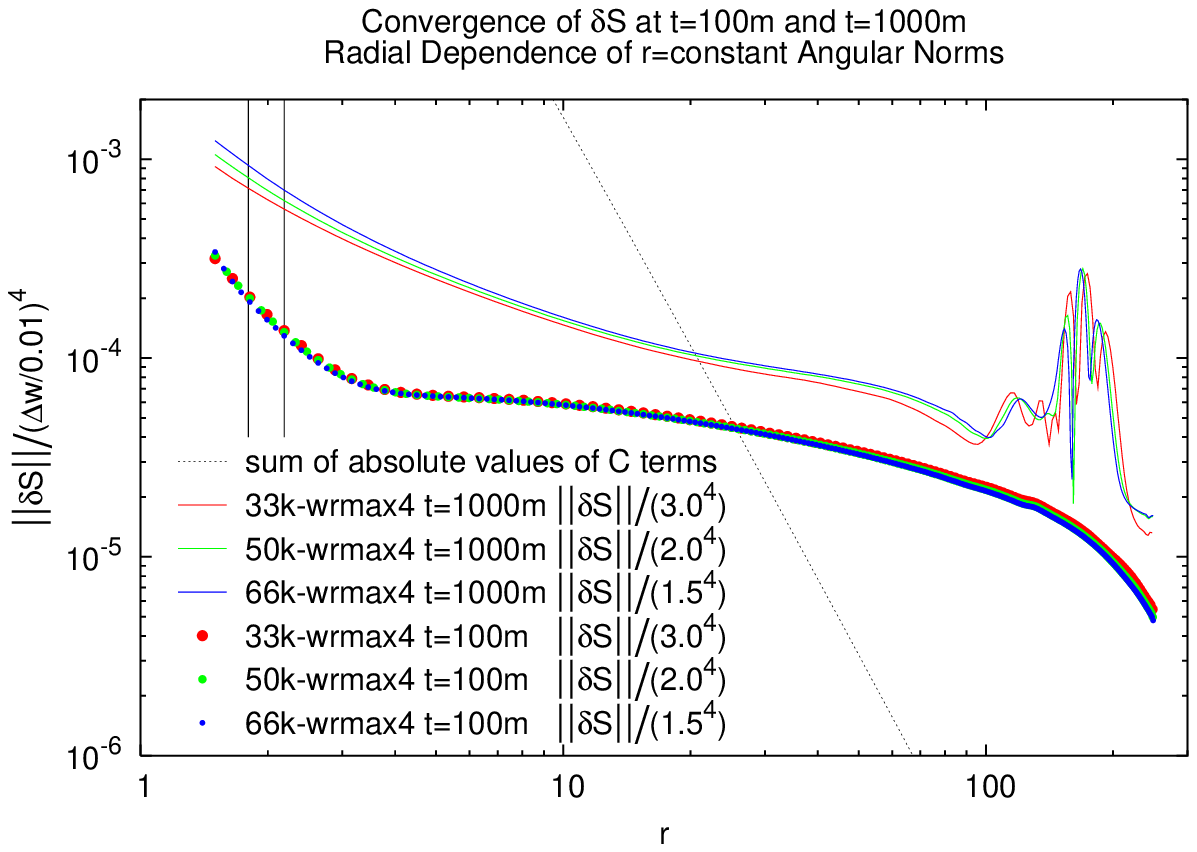}
\end{center}
\caption[Convergence of State Vector Error at $t=100m$ and $t=1000m$,
	 Radial Dependence of $r=\constant$ Angular Norms]
	{
	This figure shows the convergence of various $r=\constant$
	angular RMS-norms of the state vector error~$\delta S$
	at $t=100m$ and $t = 1000m$
	for the 33k-wrmax4, 50k-wrmax4, and 66k-wrmax4 models.
	The curves show angular RMS-norms of the state vector error
	(scaled by the 4th~power of the resolution)
	over all grid points in each $r=\constant$ shell;
	in all cases $\infty$-norms are within a factor of~$3$ of
	the RMS-norms shown.
	The diagonal dashed line shows $\| C_\absexact \|$.
	The two vertical lines show the horizon position
	and the $\wr = 0.12$ ($r = 2.19m$) position used for
	the angular convergence plots.
	}
\label{fig-converge-delta-sv-rdat-t=100+1000-wrmax=4}
\end{figure}

Figures~\ref{fig-converge-delta-sv-adat-z-t=1000-wrmax=4}
and~\ref{fig-converge-delta-sv-rdat-t=100+1000-wrmax=4}
show typical convergence results for the state vector error $\delta S$
at the same time in these same evolutions.
The convergence isn't as good as for $C$, but it's still between 3rd
and 4th~order everywhere in the grid; the overall best-fitting convergence
exponent is about~$3.8$.

\subsection{Time Dependence}

\begin{figure}[tbp]
\begin{center}
\includegraphics[width=125mm]
   {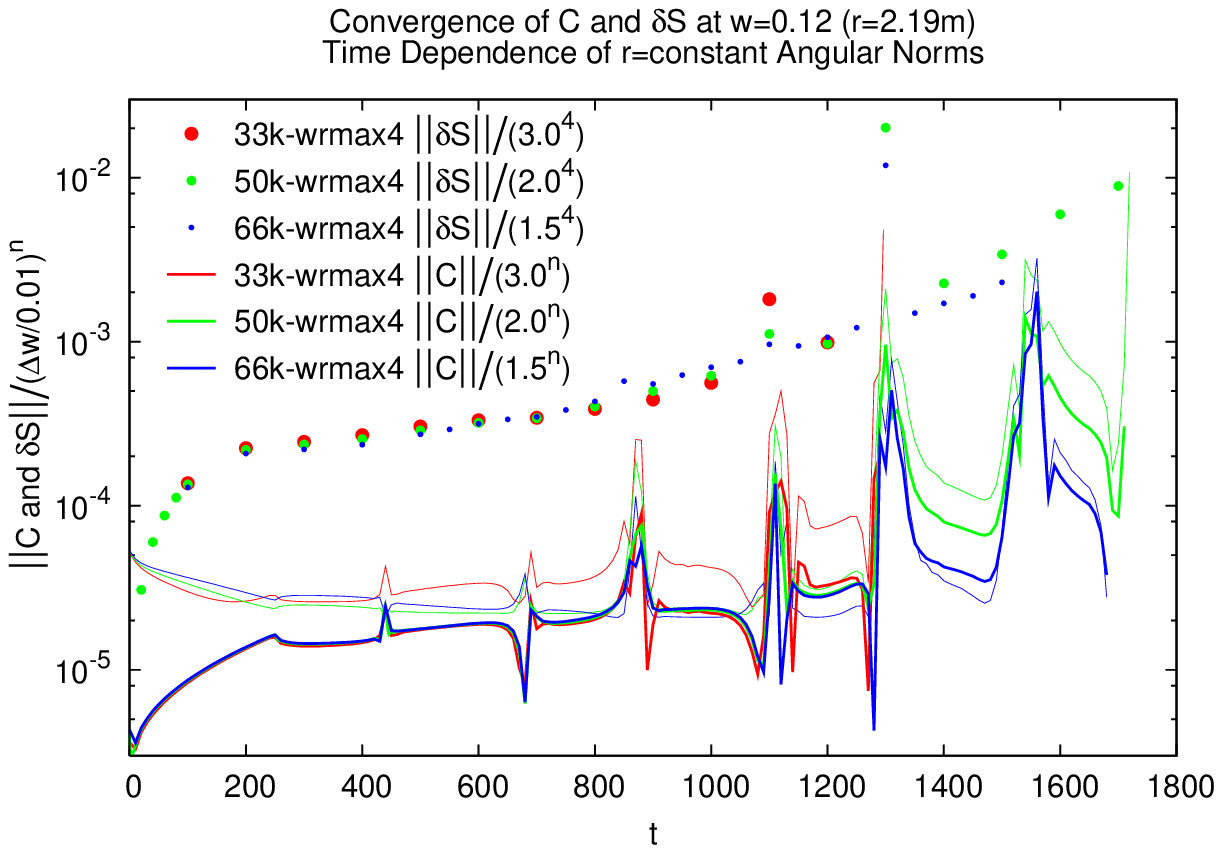}
\end{center}
\caption[Time Dependence of Convergence of Energy Constraint
	 and State Vector Error]
	{
	This figure shows the convergence of various $r=\constant$
	angular RMS-norms of the energy constraint~$C$ and the
	state vector error~$\delta S$ at the fixed radial position
	$\wr = 0.12$ ($r = 2.19m$, as a function of time,
	for the 33k-wrmax4, 50k-wrmax4, and 66k-wrmax4 models.
	The points show angular RMS-norms of the state vector error
	(scaled by the 4th~power of the resolution) over all grid points
	in each $r=\constant$ shell.
	The thick [thin] lines show angular RMS-norms of the
	energy constraint (scaled by the 4th~[3rd]~power of the resolution)
	over all patch-interior [interpatch-boundary] grid points
	in each $r=\constant$ shell.
	}
\label{fig-converge-C+delta-sv-tdat-wr=0.12}
\end{figure}

Figure~\ref{fig-converge-C+delta-sv-tdat-wr=0.12} shows the convergence
of the energy constraint~$C$ and the state vector error~$\delta S$, at
the fixed radial position $\wr = 0.12$ ($r = 2.19m$), as a function of
time.  After an initial transient lasting about $250m$, the evolution
appears to settle down to an almost-stationary state (apart from the
error waves described below) for some time.  Both $C$ and $\delta S$
generally grow with time, but remain quite small in magnitude ($\ll 1$)
throughout the first $t=1000m$ of the evolution.  ($\delta S$ seems to
grow at a faster rate than $C$; this seems to be a gauge instability
(\citet{Alcubierre94b,Alcubierre97b}.)

Notice that although the energy constraint~$C$ is substantially larger
at the interpatch boundaries than in the patch interiors (this is
visible in figure~\ref{fig-converge-C-adat-z-t=1000-wrmax=4}), this
difference doesn't grow with time, in fact it even decreases a bit.
This suggests that the the interpatch interpolation is not introducing
any instability into the evolution.

\begin{figure}[tbp]
\begin{center}
\begin{picture}(160,120)
\put(0,60){\includegraphics[width=80mm]
   {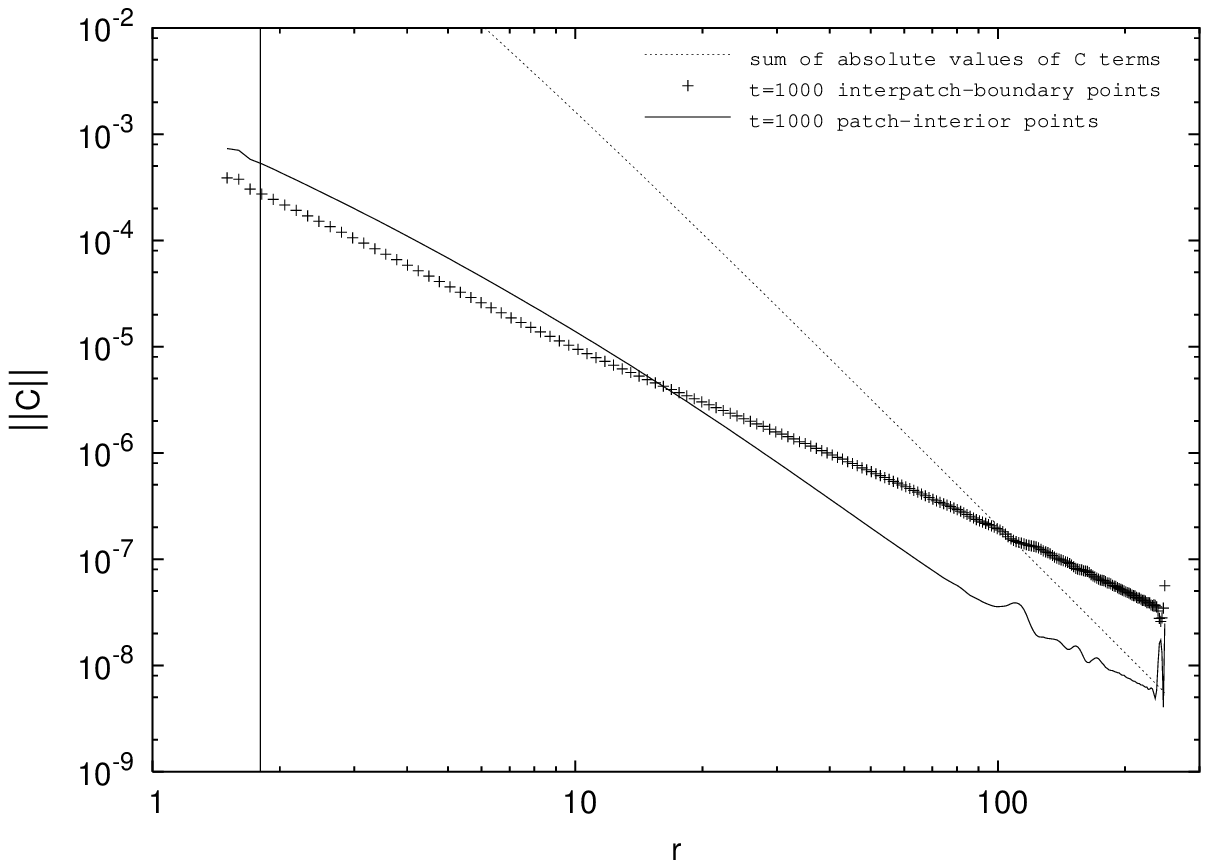}}
\put(80,60){\includegraphics[width=80mm]
   {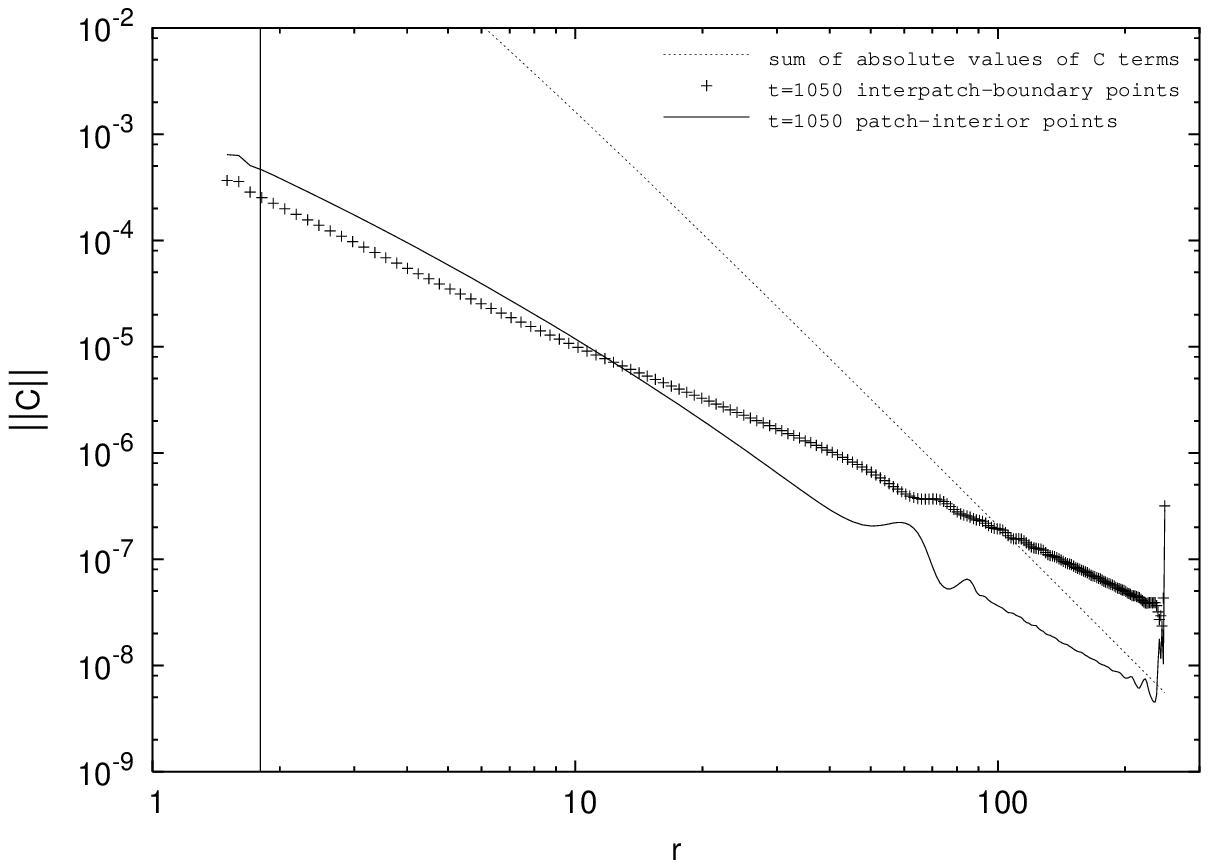}}
\put(0,0){\includegraphics[width=80mm]
   {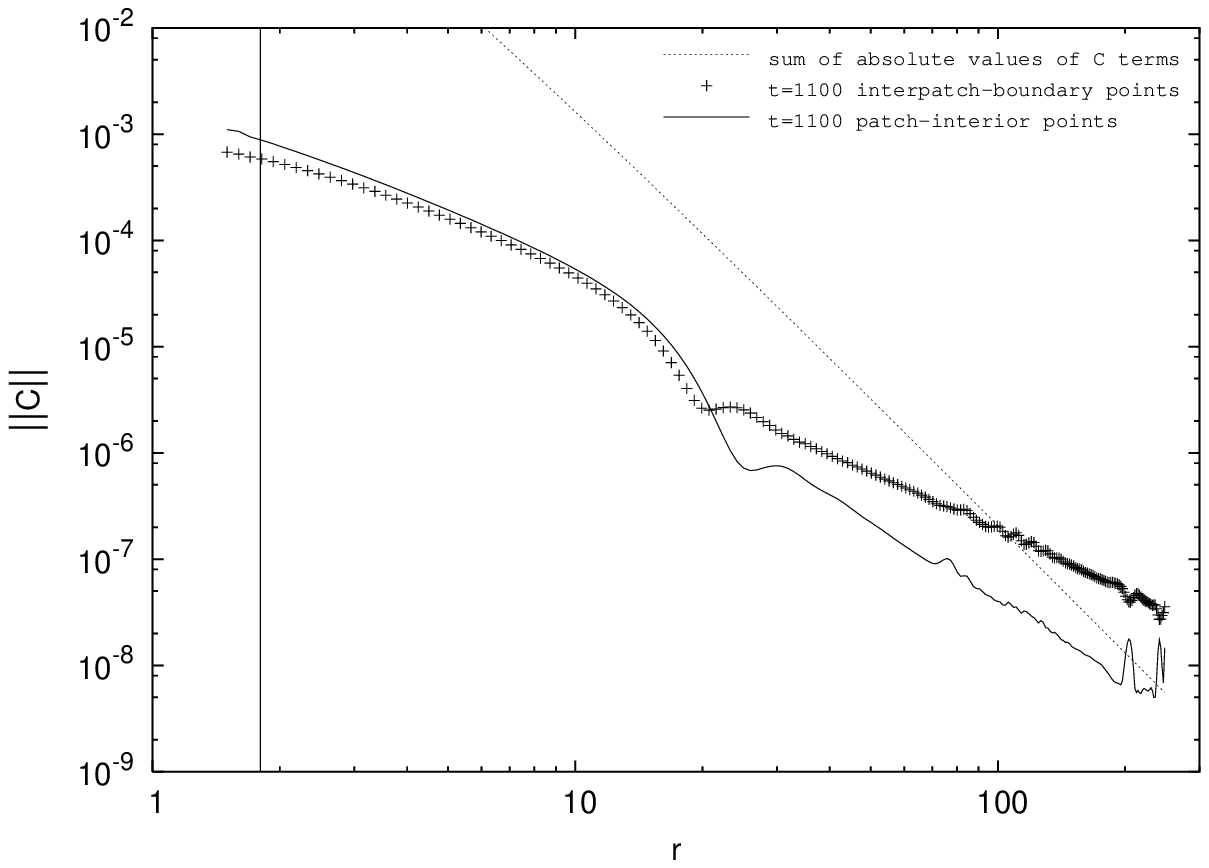}}
\put(80,0){\includegraphics[width=80mm]
   {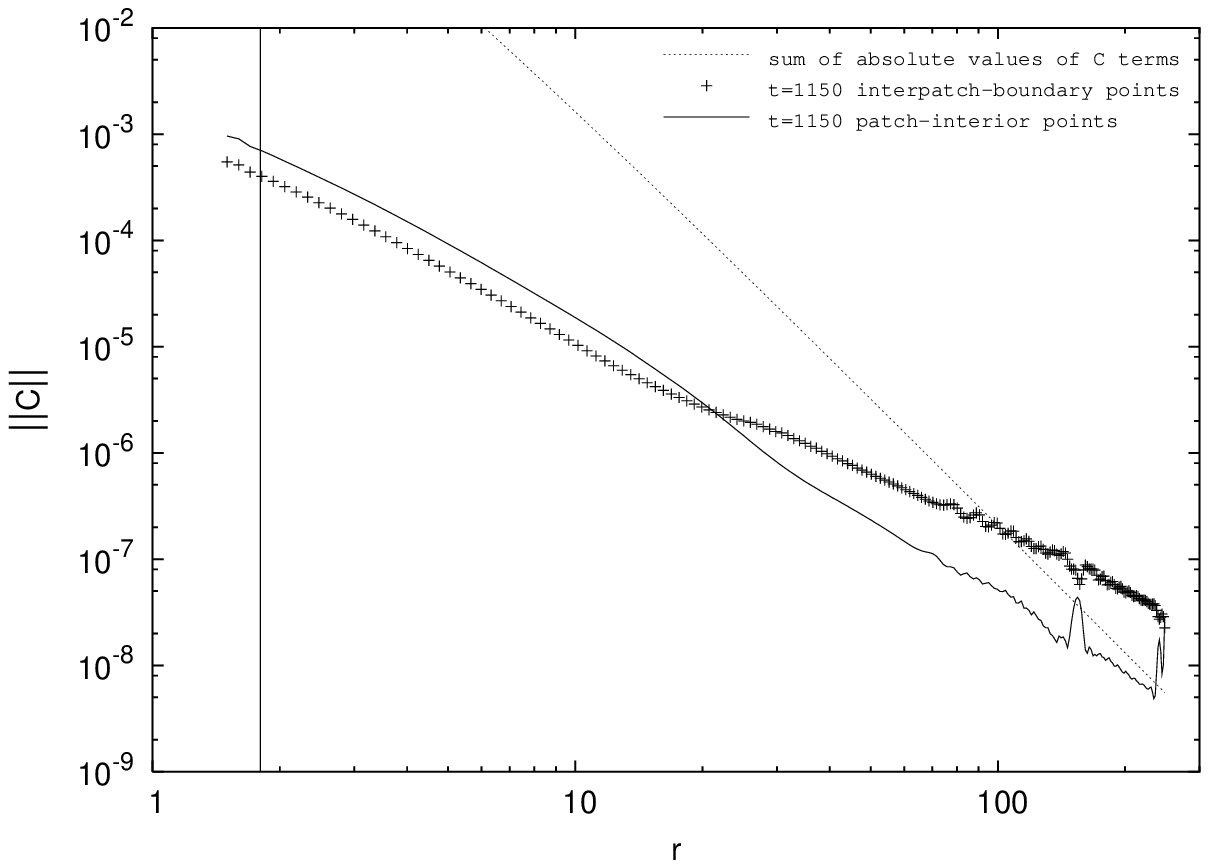}}
\end{picture}
\end{center}
\caption[Selected Frames from Movie of 50k-wrmax4 Energy Constraint]
	{
	This figure shows 4 frames from a movie showing
	the time evolution of angular RMS-norms of the
	energy constraint~$C$ for the 50k-wrmax4 model.
	The solid lines (points) show angular RMS-norms over 
	the patch-interior (interpatch-boundary)
	grid points in each $r = \constant$ shell; in all cases
	$\infty$-norms are within an order of magnitude of the
	RMS-norms shown.
	In each subplot,
	the diagonal dashed line shows $\| C_\absexact \|$
	and the vertical line shows the horizon position.
	At $t=1000m$ an inwards-propagating error wave is visible
	at $r \approx 100m$; by $t=1050m$ it has propagated in
	to $r \approx 50m$.  At $t=1100m$ the wave is just falling
	into the black hole, and at $t=1150m$ the wave has
	almost completely vanished.
	The full movie is available from
	\url{http://www.aei.mpg.de/~jthorn/research/mpe/movies/},
	and as supplemental information for this article
	on the Classical and Quantum Gravity web site.
	}
\label{fig-50k-wrmax4-C-rdat-movie}
\end{figure}

A major exception to the ``roughly stationary state'' in
figure~\ref{fig-converge-C+delta-sv-tdat-wr=0.12} is the sharp spikes
in $C$ at roughly $t=440m$, $680m$, $880m$, and succeeding times.
Figure~\ref{fig-50k-wrmax4-C-rdat-movie} shows that each ``spike''
is actually the passage of an error wave which originates at the
outer boundary and propagates inwards at approximately the speed
of light.  (The error waves are spaced roughly $250m$ apart, matching
the speed-of-light propagation time from the outer boundary inwards
to the black hole.)  At late times there are also visible error waves
propagating outwards from the strong-field region, and eventually
high-frequency oscillations in $C$ appear in the outer part of the grid.

\subsection{Overlapping versus Just-Touching Patches}

As discussed in section~\ref{sect-multipatch/synchronizing}, in my
numerical scheme adjacent patches' nominal grids must either overlap,
just touch, or be separated by a gap of one grid spacing.

\begin{figure}[tbp]
\begin{center}
\includegraphics[width=125mm]
   {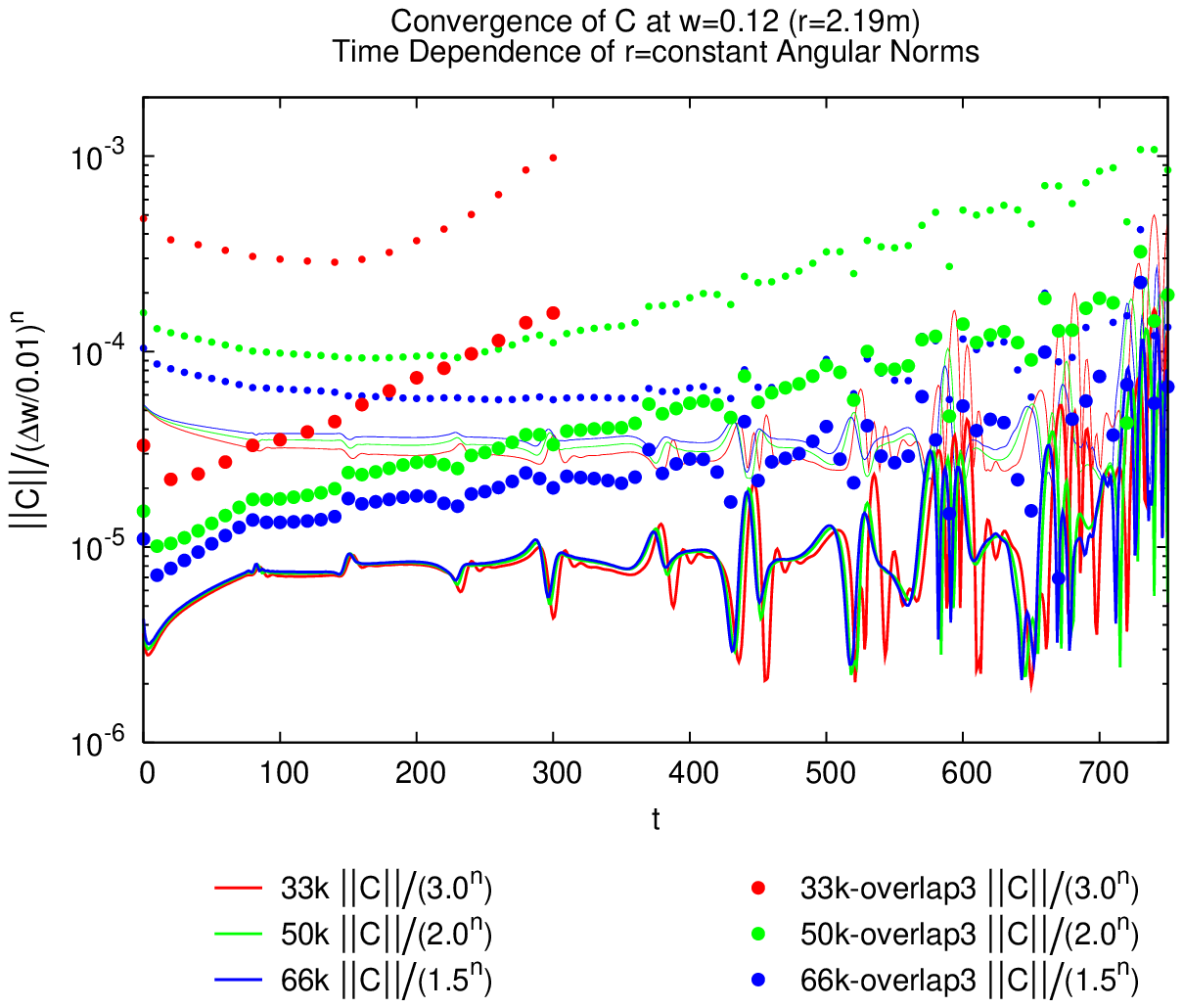}
\end{center}
\caption[Time Dependence of Convergence of Energy Constraint
	 for Touching versus Overlapping Patches]
	{
	This figure shows the convergence of various $r=\constant$
	angular RMS-norms of the energy constraint~$C$ 
	at the fixed radial position $\wr = 0.12$ ($r = 2.19m$,
	as a function of time, for the 33k, 50k, and 66k models, compared to
	the 33k-overlap3, 50k-overlap3, and 66k-overlap3 models.
	The thick [thin] lines and large [small] points
	show angular RMS-norms of the the energy constraint
	(scaled by the 4th~[3rd]~power of the resolution)
	over all patch-interior [interpatch-boundary] grid points
	in each $r=\constant$ shell.
	}
\label{fig-converge-C-overlap=3-tdat-wr=0.12}
\end{figure}

Figure~\ref{fig-converge-C-overlap=3-tdat-wr=0.12} shows a comparison
of results from models with adjacent patches just touching, versus
models with adjacent patches overlapping by $\pm 3$~grid points
about their mean nominal-grid-boundary position (\ie{} models
with adjacent patches having 7~grid points in common in their
perpendicular-to-the-boundary direction).
\footnote{
	 The results for the models with adjacent patches
	 overlapped by a fixed angular distance of
	 $\pm 9^\degree$ are generally similar to those
	 for the models with adjacent patches overlapped
	 by $\pm 3$~grid points.
	 }
{}  The just-touching models
are considerably more stable (strictly speaking, less unstable)
than the overlapping-patch models.  In particular, notice that
(apart from error waves similar to those discussed in the previous
section)
\footnote{
	 The error wave spacing again roughly matches
	 the speed-of-light propagation time from the
	 outer boundary inwards to the black hole.
	 }
{} the just-touching models display roughly constant~$C$ for
roughly~$500m$, while the overlapping-patch models display steadily
rising~$C$ in the patch interiors over the same time span.

Another interesting pattern visible in table~\ref{tab-grid-parameters}
and figure~\ref{fig-converge-C-overlap=3-tdat-wr=0.12}, is that among
the overlapping-patch models, the higher-resolution models are more
stable (less unstable).  It's hard to draw any conclusions about a
continuum limit, though, because while the just-touching models show
good 4th~(3rd)~order convergence in the patch interiors (interpatch
boundaries), even the highest-resolution overlapping-patch models
aren't yet in the asymptotic convergence regime.
\footnote{
	 The formal convergence exponents for the
	 highest-resolution pair of overlapping-patch
	 models are about~5(4.5) in the patch interiors
	 (interpatch boundaries).  Both of these exponents
	 are greater than~4 (the order of the finite
	 differencing scheme), so the models can't be
	 in the asymptotic convergence regime.
	 }


\section{Discussion and Conclusions}

Because excision in Cartesian grids is difficult, and polar spherical
grids suffer from $z$~axis coordinate singularities, there is growing
interest in multiple-patch black hole excision schemes which allow a
smooth $r = \constant$ excision surface.  As well as greatly simplifying
excision, such schemes also provide a smooth outer boundary, and (by
placing the $r = \constant$ shells of grid points nonuniformly in radius)
they can give high resolution close to the black hole while still having
the outer boundary relatively far away.

In this paper I present a detailed description of what I believe is the
first multiple-patch excision scheme for the full nonlinear Einstein
equations in $3+1$ dimensions, together with sample numerical results
from a prototype implementation of this scheme.  I use the BSSN form
of the $3+1$ equations.

I use a \defn{ghost zone} technique to handle the interpatch boundaries,
with all the dynamical fields interpolated into each patch's ghost zones
from adjacent patches.  I use an \defn{inflated-cube} 6-patch system
of the form $\{ r \times (\text{6 angular patches covering $S^2$}) \}$.
By suitably choosing the angular coordinates, this allows adjacent
patches to always share the angular coordinate perpendicular to their
mutual boundary, so the interpatch interpolation need only be done in
the (angular) dimension parallel to the boundary.

I use each patch's local coordinates to define its tensor basis,
so this basis is different from one patch to another, and hence the
interpatch interpolation must include a change of tensor basis.
This is straightforward for all the BSSN variables except for the
$\tilde{\Gamma}^i$.  The $\tilde{\Gamma}^i$ aren't tensors or tensor
densities, so their change-of-basis transformation law includes terms
containing spatial derivatives of the BSSN conformal factor~$\cphi$.
Computing these derivatives numerically is difficult because this
must be done in the ghost zones.  I currently solve this problem by
using a ghost zone twice as wide for the BSSN conformal factor~$\cphi$
as for the other BSSN field variables.  This works, but is somewhat
cumbersome; it would be interesting to explore the alternative scheme
I describe (but have not implemented) where the ghost zones can be
the same width for all the field variables.

I have implemented my multiple-patch scheme in a prototype numerical
code, using 4th~order finite differencing in space and time, and
5th~order Lagrange polynomial interpatch interpolation.  I use
ghost zones which are 4~points wide for the BSSN conformal
factor~$\cphi$, and 2~points wide for all the other dynamical
field variables.  My finite differencing scheme is a straightforward
generalization to 3-D of the 1-D finite differencing scheme I described
in~\citet{Thornburg99}.  The overall accuracy of the scheme is 4th~order
in the patch interiors, and 3rd~order close to the interpatch and
radial boundaries (though this could be raised to 4th~order everywhere
by adjusting the interpolation and radial extrapolation operators).

In tests of the evolution of octant-symmetry Kerr initial data with
various numerical parameters, this scheme performs quite well, with
evolutions maintaining good convergence and preserving the energy
constraint near the black hole to better than $1$\% for $\gtsim 1000m$,
and lasting up to $\gtsim 1500m$ before crashing.  The degradation
of accuracy at later times, and the eventual crashes, appear to be
due to outer-boundary effects, {\em not\/} to any problems inherent
to the multiple-patch scheme.

Two obvious areas in which to extend these results would be to drop
the octant-symmetry assumption, and to evolve more general (non-stationary)
initial data.  I hope to experiment in these directions soon.

The numerical code I have used for these results is a standalone one,
not integrated into any standard numerical-relativity toolkit such as
Cactus (\citet{Goodale02a}).  Such an integration will pose some
software-engineering problems, but should have large benefits for
furthering the practical use of multiple-patch schemes.

How could my multiple-patch scheme be extended to handle multiple
(moving) black holes?  The easiest way appears to be to use an
inflated-cube set of patches around (and kept dynamically centered on)
each (excised) black hole, a Cartesian patch or patches filling
the space between the black holes, and probably an outer set of
inflated-cube patches outside both black holes.
\footnote{
	 Using a Cartesian patch or patches between
	 the black holes avoids the $r=0$ singularities
	 of the outer set of inflated-cube patches.
	 }
$^,$
\footnote{
	 Such a scheme is essentially a 3-D inflated-cube
	 analog of the axisymmetric multiple-patch scheme
	 I have previously used for constructing
	 2-black-hole conformally-flat initial data
	 (\citet{Thornburg85,Thornburg87}).
	 }
{}  This scheme requires multidimensional interpatch interpolation
between the inflated-cube and Cartesian patches, but the experience
of \citet{Calabrese2003:excision-and-summation-by-parts}, as well
as the extensive experience with similar schemes in computational
fluid dynamics (see references cited in section~\ref{sect-introduction}),
suggests that this probably doesn't cause any significant problems.
Such schemes should be a fruitful area for further research.

\appendix

\section{Details of the Ghost-Zone Synchronization Algorithm}
\label{app-ghost-zone-sync-details}

In this appendix I give a detailed description of my ghost-zone
synchronization algorithm.

Figure~\ref{fig-ghost-zone-sync-algorithm} shows the major steps of
the algorithm.  The main complexity of the algorithm
(steps~\ref{step-symmetry-non-corners} to~\ref{step-symmetry-corners})
is in the handling of the angular ghost zones.  Because this is done
independently in each $r=\constant$ angular shell of grid points, for
the rest of this appendix I drop the $r$ coordinate and just describe
the angular computations.

\begin{figure}[tbp]
\begin{enumerate}
\item	\label{step-radial-extrap-1}
	Radially extrapolate the grid function(s) in the nominal grid
	to compute their values in the nominal--angular-grid part of
	each patch's inner and outer (radial) ghost zones.
\item	\label{step-symmetry-non-corners}
	Apply the appropriate symmetry mappings to compute grid function
	values in the non-angular-corner nominal-radial-grid part of
	each symmetry ghost zone.
\item	\label{step-interpatch-interp}
	Use interpatch interpolation to compute grid function values
	in the nominal-radial-grid part of each interpatch ghost zone.
\item	\label{step-change-of-basis}
	Do a change-of-basis transformation on each interpatch-interpolated
	value.  (See section~\ref{sect-multipatch/change-of-tensor-basis}
	for details of this.)
\item	\label{step-symmetry-corners}
	Apply the appropriate symmetry mapping to compute grid function
	values in the angular corners of the nominal-radial-grid part
	of each symmetry ghost zone.
\item	\label{step-radial-extrap-2}
	Radially extrapolate the grid function(s) in the nominal grid
	to compute their values in the angular-ghost-zones part of
	each patch's inner and outer (radial) ghost zones.
\end{enumerate}
\caption[Ghost-Zone Synchronization Algorithm]
	{
	This figure shows my algorithm for synchronizing ghost zones.  
	}
\label{fig-ghost-zone-sync-algorithm}
\end{figure}

Consider an interpatch boundary between patches~$\p$ and~$\q$, and
define $(\rho,\sigma)$~angular coordinates relative to this boundary,
$\perp$~perpendicular to the boundary and (in each patch)
$\parallel$~parallel to the boundary.
(Recall from section~\ref{sect-multipatch/choice-of-patches+coords}
that the patches share a common $\perp$ coordinate, while each has
its own $\parallel$ coordinate.)  Corresponding to $(\perp,\parallel)$,
I define $(\iperp,\ipar)$ as integer grid coordinates to describe
the synchronization algorithm, and also as integer array indices for
subscripting grid functions.

\begin{figure}[tbp]
\begin{center}
\begin{picture}(160,160)
\put(38,110){
	    \begin{picture}(0,0)
	    \put(-5,40){Part~(a)}
	    \put(-34.00,-19.75){\includegraphics{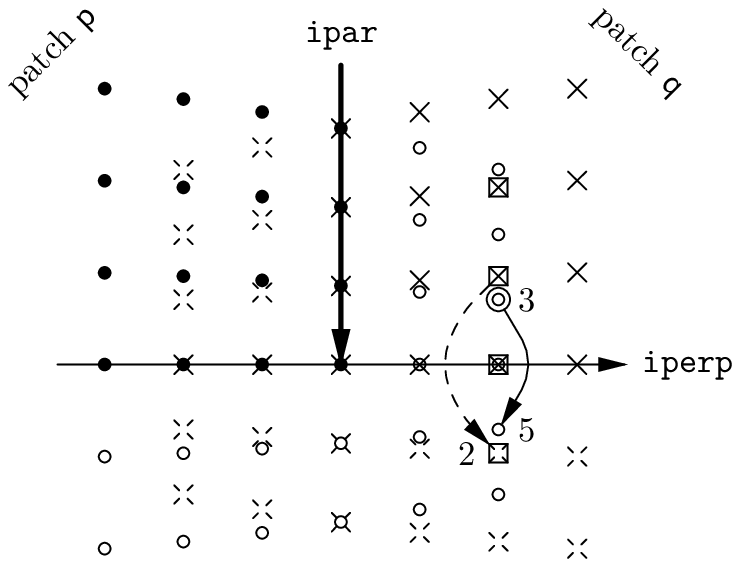}}
	    \end{picture}
	    }
\put(88,75){
	   \begin{picture}(0,0)
	   \put(0,0){\includegraphics{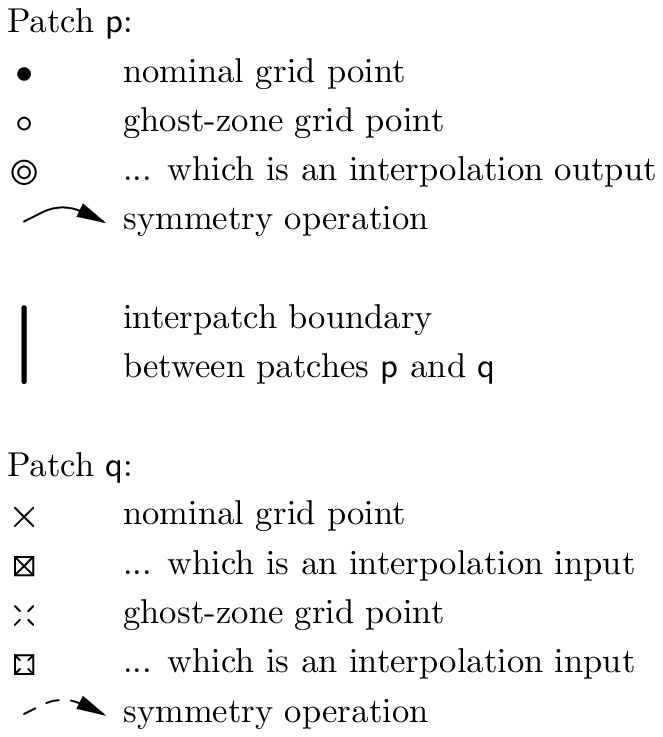}}
	   \end{picture}
	   }
\put(45,30){
	   \begin{picture}(0,0)
	   \put(-5,50){Part~(b)}
	   \put(-40.50,-33.70){\includegraphics{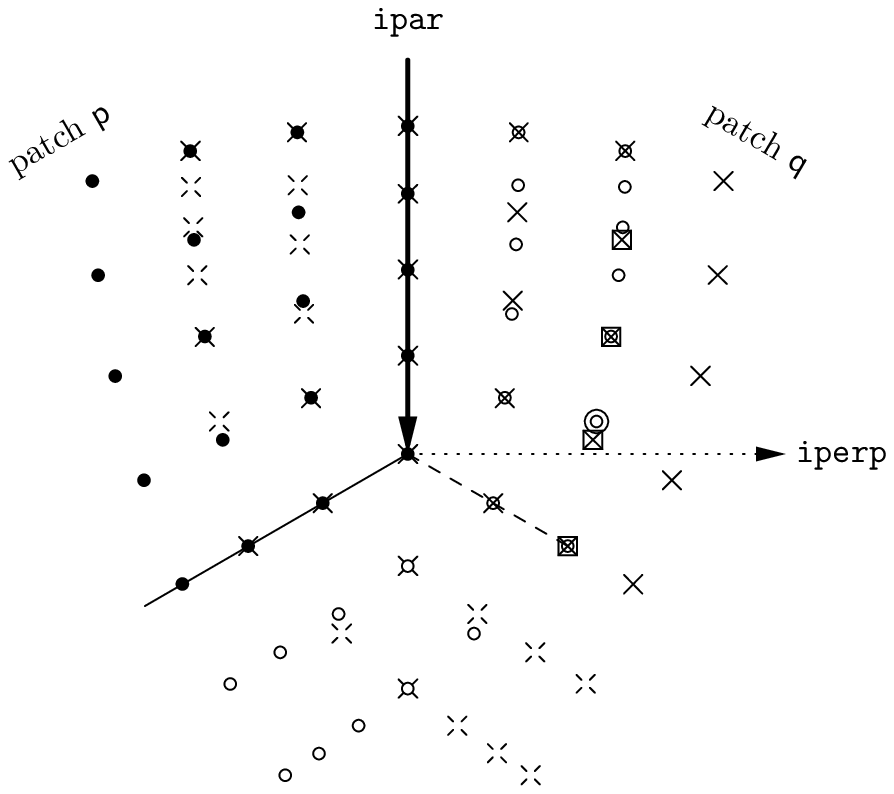}}
	   \end{picture}
	   }
\end{picture}
\end{center}
\caption[Interpatch Interpolation Details]
	{
	This figure shows details of the interpatch interpolation
	process for ghost-zone synchronization.
	The interpatch boundary between patches~$\p$ and~$\q$
	is shown as a heavy vertical line in each part of the figure.
	In both parts of the figure, $\iperp$~increases to the right
	and $\ipar$~increases downwards, though these coordinates
	aren't orthogonal.
	Part~(a) shows the case where $\p$'s adjacent ghost zone
	on the maximum-$\ipar$ end is a symmetry ghost zone
	(for this example, one corresponding to a reflection
	 symmetry across the $\iperp$ axis).
	Sample output points from various steps of the
	ghost-zone synchronization algorithm are labelled.
	Part~(b) shows the case where $\p$'s adjacent ghost zone
	on the maximum-$\ipar$ end is an interpatch ghost zone.
	The arrows show symmetry operations,
	the boxed points are interpatch interpolation inputs, and
	the circled point is an interpatch interpolation output.
	For the case shown here (a 4-point interpatch interpolation
	molecule), the interpolation can be kept centered, but
	for a larger molecule it must be off-centered.
	}
\label{fig-interpatch-interp-details}
\end{figure}

Since these coordinates are patch-dependent, I use a notation inspired
by the \Cplusplus{} programming language, where $(\p.\iperp,\p.\ipar)$
refers to the $(\iperp,\ipar)$~coordinates of patch~$\p$ relative to
this boundary.  Figure~\ref{fig-interpatch-interp-details} illustrates
the geometry of the patches, ghost zones, and coordinates.

Suppose $\p$'s~nominal grid is the rectangle
\begin{equation}
(\iperp,\p.\ipar)
	\in [\p.\iperp_\nominalmin, \p.\iperp_\nominalmax]
	    \times
	    [\p.\ipar_\nominalmin, \p.\ipar_\nominalmax]
\end{equation}
and similarly for~$\q$.  Consider one of $\p$'s~(angular)
ghost zones~$\gz$, which overlaps~$\q$'s nominal grid (that is,
patch-$\p$ grid function values in $\gz$ are to be obtained by
interpolating from patch~$\q$).  Without loss of generality suppose
that the coordinates are oriented such that~$\gz$ lies on $\p$'s
maximum-$\iperp$ boundary, so $\gz$'s~extent in the $\iperp$ direction
is $\iperp \in [\p.\iperp_\nominalmax+1, \p.\iperp_\nominalmax+\gzw]$,
where $\gzw$ is the ghost zone width.
\footnote{
	 $\gzw = 2$ for all the numerical results presented here,
	 and for the example in figure~\ref{fig-interpatch-interp-details}.
	 }

I define $\gz$'s extent in the $\p.\ipar$~direction to depend on the
type (symmetry or interpatch) of $\p$'s adjacent ghost zones on $\gz$'s
minimum-$\p.\ipar$ and maximum-$\p.\ipar$ ends:
\begin{itemize}
\item	If $\p$'s~adjacent ghost zone on an end is a symmetry ghost zone,
	then I define $\gz$ to {\em not\/} include their mutual corner.
	That is, for example, if $\p$'s maximum-$\p.\ipar$~ghost zone
	is a symmetry ghost zone, I define $\gz$'s maximum-$\p.\ipar$~extent
	to be $\p.\ipar \leq \p.\ipar_\nominalmax$.  Part~(a) of
	figure~\ref{fig-interpatch-interp-details} illustrates this case.
\item	If $\p$'s~adjacent ghost zone on an end is an interpatch
	ghost zone, then I define $\gz$ to include up to the
	diagonal line of their mutual corner, arbitrarily breaking ties
	by including the diagonal line itself into the $\rho$~ghost zone.
	That is, for example, if $\p$'s maximum-$\p.\ipar$~ghost zone is an
	interpatch ghost zone, I define $\gz$'s maximum-$\p.\ipar$~extent
	at each $\iperp$ to be
	\begin{equation}
	\p.\ipar \leq
		\p.\ipar_\nominalmax
		+ (\iperp - \p.\iperp_\nominalmax)
		- \left\{
		  \begin{array}{ll}
		  0	& \text{if $\perp$ is $\p$'s $\rho$ coordinate}	\\
		  1	& \text{if $\perp$ is $\p$'s $\sigma$ coordinate}
		  \end{array}
		  \right.
	\end{equation}
	Part~(b) of figure~\ref{fig-interpatch-interp-details}
	illustrates this case, with the diagonal line shown dashed.
\end{itemize}

Next, I define $\gz$'s \defn{patch interpolation region}~$\R$ in~$\q$:
This is the set of $\q$~grid points from which $\gz$'s interpatch
interpolation will use data.  Since the interpolation is only done
in 1-D along $\q.\ipar$~lines
(these run vertically in figure~\ref{fig-interpatch-interp-details}),
$\R$ has the same extent in the $\iperp$~direction
(horizontally in figure~\ref{fig-interpatch-interp-details})
as $\gz$.  I define $\R$'s~extent in the $\q.\ipar$~direction
to depend on the type (symmetry or interpatch) of $\q$'s adjacent
ghost zones on each of $\q$'s minimum-$\ipar$ and maximum-$\ipar$ ends:
\begin{itemize}
\item	If $\q$'s adjacent ghost zone on an end is a symmetry
	ghost zone, then I allow $\gz$'s interpolation to use data from
	that symmetry ghost zone (as well as from $\q$'s nominal grid).
\footnote{
	 Since $\gz$ is contained in $\q$'s nominal grid 
	 in the $\iperp$ direction, these symmetry-ghost-zone
	 points have already been synchronized in
	 step~\ref{step-symmetry-non-corners} of the
	 synchronization algorithm of
	 figure~\ref{fig-ghost-zone-sync-algorithm},
	 so it's ok to use their values here (in
	 step~\ref{step-interpatch-interp}).
	 }
{}	That is, for example, if $\q$'s maximum-$\q.\ipar$ ghost zone
	is a symmetry ghost zone, I define $\R$'s maximum-$\q.\ipar$
	extent to be $\q.\ipar \le \q.\ipar_\nominalmax + \gzw$.
	So long as the ghost zone is wide enough, this allows
	$\gz$'s~interpolations to be kept centered even at $\gz$~points
	within an interpolation-molecule-radius of a symmetry boundary.
	Part~(a) of figure~\ref{fig-interpatch-interp-details}
	illustrates this case.
\item	If $\q$'s adjacent ghost zone on an end is an interpatch
	ghost zone (this can only happen at a \defn{triple corner}
	where 3~patches meet), then at present I only allow
	$\gz$'s~interpolation to use data from $\q$'s nominal grid
	on that end.  That is, for example, if $\q$'s maximum-$\q.\ipar$
	ghost zone is an interpatch ghost zone, I define $\R$'s
	maximum-$\q.\ipar$ extent to be $\q.\ipar \le \q.\ipar_\nominalmax$.
	Part~(b) of figure~\ref{fig-interpatch-interp-details}
	illustrates this case.  Notice that for a 4-point or smaller
	interpatch interpolation molecule, the interpatch interpolation
	can always be kept centered, while for a 5-point or larger
	molecule, the interpolation must be off-centered for points
	close to the triple corner.

	If interpatch ghost zones are interpolated in some sequential
	order in step~\ref{step-interpatch-interp} of the
	synchronization algorithm of
	figure~\ref{fig-ghost-zone-sync-algorithm}, it would be
	possible to keep more of the interpolations centered by
	expanding $\R$ to include points of the adjacent interpatch
	ghost zone if that ghost zone came earlier than~$\gz$ in the
	sequential order, and thus had already been interpolated.
	I haven't investigated this possibility in detail, but
	(by keeping more of the interpolations centered) it might
	well lead to better numerical stability in the time evolution.
\end{itemize}

\section{\jtbold Transformation Properties of the $\tilde{\Gamma}^i$}
\label{app-Gamma-tilde-u-cxform}

In this appendix I outline the derivation of the interpatch
coordinate-transformation law~\eqref{eqn-Gamma-tilde-u-cxform}
for the BSSN conformal Christoffel symbols $\tilde{\Gamma}^i$.

The transformation law for the non-conformal Christoffel symbols
$\Gamma^i$ is well-known (\citet[equation~$(10.26)$]{Misner73}):
\begin{equation}
\Gamma^a(\p)
	= X^a{}_i \Gamma^i(\q) + X^a{}_i Y^i{}_{bc} g^{bc}(\p)
						\label{eqn-Gamma-u-cxform}
\end{equation}
Substituting the identity (valid so long as we're in a 3-D slice)
\begin{equation}
\tilde{\Gamma}^a
	= e^{4\cphi} \Gamma^a + 2 \tilde{g}^{ab} \partial_b \cphi
					\label{eqn-Gamma-tilde-u=fn(Gamma-u)}
\end{equation}
into each side of this, then using the $\cphi$ transformation
law~\eqref{eqn-cphi-cxform} and simplifying, gives the final
$\tilde{\Gamma}^i$ transformation law~\eqref{eqn-Gamma-tilde-u-cxform}.


\section{Outer Boundary Conditions}
\label{app-outer-BCs}

Basically, I use Sommerfeld outer boundary conditions, applied to each
coordinate component of the field variables.  However, there's a
complication: because the $(r,\rho,\sigma)$ tensor basis isn't
asymptotically Cartesian (for example, asymptotically there are
$r^2$~factors in the angular metric components), I need to transform
to the locally Cartesian basis $\{ \bar{x}^i \}$ before applying the
usual Sommerfeld condition.  In this appendix I use an overbar to
denote components in a locally Cartesian basis, as in $\bar{f}$.

For each field variable~$f$, I actually apply a Sommerfeld condition
not to $\bar{f}$~itself, but rather to the deviation
$\delta\bar{f} \equiv \bar{f} - \bar{f}_\background$ of~$\bar{f}$
from its (time-independent) value in a specified background slice.
The Sommerfield condition for $f$ is thus
$\partial_t (r^n \bar{f}) = - c \partial_r (r^n \delta\bar{f})$,
where $c$ is the propagation speed and $n$ the falloff exponent,
or equivalently
\begin{equation}
\partial_t \bar{f}
	= -c \left[
	     \partial_r \delta\bar{f} + \frac{n}{r} \delta\bar{f}
	     \right]
							\label{eqn-Sommerfeld}
\end{equation}

For 3-scalars such as $\alpha$ and $K$ I can apply this boundary condition
directly.  The other BSSN variables need to be transformed to locally
Cartesian coordinates for the Sommerfeld condition~\eqref{eqn-Sommerfeld}
to be applied.  Proceeding analogously to
section~\ref{sect-multipatch/change-of-tensor-basis},
I define the transformation matrices
\begin{subequations}
\begin{eqnarray}
\bar{X}^a{}_i
	& = &	\ddfrac{\partial x^a}{\partial \bar{x}^i}		\\
\bar{Y}^i{}_a
	& = &	\ddfrac{\partial \bar{x}^i}{\partial x^a}		\\
\bar{Y}^i{}_{ab}
	& = &	\ddfrac{\partial^2 \bar{x}^i}{\partial x^a \partial x^b}
\end{eqnarray}
\end{subequations}
These are straightforward to compute analytically from the coordinate
definitions~\eqref{eqn-mu-nu-phi}.  Note that the $\bar{X}^a{}_i$ and
$\bar{Y}^i{}_a$ matrices are inverses.

The BSSN conformal factor~$\cphi$ transforms as
\begin{equation}
\bar{\cphi} = \cphi - \tfrac{1}{6} \log |\bar{Y}|
\end{equation}
where $\bar{Y} \equiv \det \left[ \bar{Y}^i{}_a \right]$.  Because
$Y \equiv Y(x^a)$ is a known coefficient, not a dynamical variable,
$\delta\bar{\cphi} = \delta\cphi$, so for purposes of the Sommerfeld
condition~\eqref{eqn-Sommerfeld} I can treat $\cphi$ as a 3-scalar,
and apply~\eqref{eqn-Sommerfeld} directly.

The BSSN $\tilde{g}_{ij}$ and $\tilde{A}_{ij}$ transform as
\begin{equation}
\bar{\tilde{S}}_{ij}
	= |\bar{Y}|^{2/3} \bar{X}^a{}_i \bar{X}^b{}_j \tilde{S}_{ab}
\end{equation}
where $\tilde{S}_{ij} \in \{ \tilde{g}_{ij}, \tilde{A}_{ij} \}$.
Substituting this into both sides of the Sommerfeld
condition~\eqref{eqn-Sommerfeld}, multiplying both sides by
$|\bar{Y}|^{-2/3} \bar{Y}^i{}_c \bar{Y}^j{}_d$ and summing over $i$ and $j$,
and simplifying, then gives the boundary condition
\begin{equation}
\partial_t \tilde{S}_{ij}
	= -c \left[
	     \partial_r \delta\tilde{S}_{ij}
	     + \frac{n}{r} \delta\tilde{S}_{ij}
	     + \bar{Z}^k{}_i \delta\tilde{S}_{kj}
	     + \bar{Z}^k{}_j \delta\tilde{S}_{ki}
	     + \tfrac{2}{3} (\partial_r \log |\bar{Y}|) \delta\tilde{S}_{ij}
	     \right]
							\label{eqn-Sdd-outer-BC}
\end{equation}
where $\bar{Z}^a{}_b \equiv \bar{Y}^i{}_b \partial_r \bar{X}^a{}_i$.

The BSSN $\tilde{\Gamma}^i$ transform in a more complicated way:
Applying the interpatch transformation law~\eqref{eqn-Gamma-u-cxform}
to the coordinates $x^a(\p)$ and $\bar{x}^i(\q)$, and using the
identity~\eqref{eqn-Gamma-tilde-u=fn(Gamma-u)} and its inverse
in each of these coordinate systems, gives the transformation law
\begin{equation}
\bar{\tilde{\Gamma}}^i
	= |\bar{Y}|^{-2/3}
	  \left[
	      \bar{Y}^i{}_a \tilde{\Gamma}^a
	  - 2 \bar{Y}^i{}_a               \tilde{g}^{ab} \partial_b \cphi
	  + 2 \bar{Y}^i{}_a \bar{Y}^j{}_b \tilde{g}^{ab} \partial_j \cphi
	  -   \bar{Y}^i{}_{ab}            \tilde{g}^{ab}
	  + \tfrac{1}{3}
              \bar{Y}^i{}_a \bar{Y}^j{}_b (\partial_j \log |\bar{Y}|)
					  \tilde{g}^{ab} 
	  \right]
\end{equation}
Substituting this into both sides of the Sommerfeld
condition~\eqref{eqn-Sommerfeld}, multiplying both sides by
$|\bar{Y}|^{2/3} \bar{X}^c{}_i$ and summing over $i$, and simplifying,
then gives the boundary condition
\begin{eqnarray}
\partial_t \tilde{\Gamma}^c
	& = &	2 \left(
		  \partial_t \tilde{g}^{bc} \partial_b \cphi
		  + \tilde{g}^{bc} \partial_b \partial_t \cphi
		  \right)
		- 2 \bar{Y}^j{}_b
		    \left(
		      \partial_t \tilde{g}^{bc} \partial_j \cphi
		    + \tilde{g}^{bc} \partial_j \partial_t \cphi
		    \right)
							\nonumber	\\
	&   &	{}
		+ \bar{X}^c{}_i \bar{Y}^i{}_{ab} \partial_t \tilde{g}^{ab}
		+ \tfrac{1}{3} \bar{Y}^j{}_b (\partial_j \log |\bar{Y}|)
		  \partial_t \tilde{g}^{bc}
							\nonumber	\\
	&   &	{}
		- c \biggl\{
		    \bar{W}^c{}_a \delta\tilde{\Gamma}^a
		    + \partial_r \delta\tilde{\Gamma}^c
		    - 2 \bar{W}^c{}_a
			\left(
			  \delta\tilde{g}^{ab} \partial_b \cphi
			+ \tilde{g}^{ab} \partial_b \delta\cphi
			\right)
							\nonumber	\\
	&   &	\quad\qquad
		    - 2 \left(
			  \partial_r \delta\tilde{g}^{bc} \partial_b \cphi
			+ \delta\tilde{g}^{bc} \partial_{rb} \cphi
			+ \partial_r \tilde{g}^{bc} \partial_b \delta\cphi
			+ \tilde{g}^{bc} \partial_{rb} \delta\cphi
			\right)
							\nonumber	\\
	&   &	\quad\qquad
		{}
		    + 2 \bar{W}^c{}_a \bar{Y}^j{}_b
			\left(
			  \delta\tilde{g}^{ab} \partial_j \cphi
			+ \tilde{g}^{ab} \partial_j \delta\cphi
			\right)
		    + 2 (\partial_r \bar{Y}^j{}_b)
			\left(
			  \delta\tilde{g}^{bc} \partial_j \cphi
			+ \tilde{g}^{bc} \partial_j \delta\cphi
			\right)
							\nonumber	\\
	&   &	\quad\qquad
		{}
		    + 2 \bar{Y}^j{}_b
			\left(
			  \partial_r \delta\tilde{g}^{bc} \partial_j \cphi
			+ \delta\tilde{g}^{bc} \partial_{rj} \cphi
			+ \partial_r \tilde{g}^{bc} \partial_j \delta\cphi
			+ \tilde{g}^{bc} \partial_{rj} \delta\cphi
			\right)
							\nonumber	\\
	&   &	\quad\qquad
		{}
		    - \bar{W}^c{}_{ab} \delta\tilde{g}^{ab}
		    - \bar{X}^c{}_i \bar{Y}^i{}_{ab}
		      \partial_r \delta\tilde{g}^{ab}
							\nonumber	\\
	&   &	\quad\qquad
		{}
		    - \tfrac{1}{3} \bar{W}^c{}_a \bar{Y}^j{}_b
		      (\partial_j \log |\bar{Y}|) \delta\tilde{g}^{ab}
		    - \tfrac{1}{3} \partial_r \bar{Y}^j{}_b
		      (\partial_j \log |\bar{Y}|) \delta\tilde{g}^{bc}
							\nonumber	\\
	&   &	\quad\qquad
		{}
		    - \tfrac{1}{3} \bar{Y}^j{}_b
		      (\partial_{rj} \log |\bar{Y}|) \delta\tilde{g}^{bc}
		    - \tfrac{1}{3} \bar{Y}^j{}_b
		      (\partial_r \log |\bar{Y}|)
		      \partial_r \delta\tilde{g}^{bc}
							\nonumber	\\
	&   &	\quad\qquad
		{}
		    + \left(
		      \frac{n}{r} - \tfrac{2}{3} \partial_r \log |\bar{Y}|
		      \right)
		      \Bigl[
		      \delta\tilde{\Gamma}^c
		      - 2 \left(
			    \delta\tilde{g}^{bc} \partial_b \cphi
			  + \tilde{g}^{bc} \partial_b \delta\cphi
			  \right)
		      + 2 \bar{Y}^j{}_b
			  \left(
			    \delta\tilde{g}^{bc} \partial_j \cphi
			  + \tilde{g}^{bc} \partial_j \delta\cphi
			  \right)
							\nonumber	\\
	&   &	\qquad\qquad\qquad\qquad\qquad\qquad\qquad
		{}
		      - \bar{X}^c{}_i \bar{Y}^i{}_{ab} \delta\tilde{g}^{ab}
		      - \tfrac{1}{3} \bar{Y}^j{}_b
			(\partial_j \log |\bar{Y}|) \delta\tilde{g}^{bc}
		      \Bigr]
		    \biggr\}
					\label{eqn-Gamma-tilde-u-outer-BC}
\end{eqnarray}
where $\bar{W}^a{}_b \equiv \bar{X}^a{}_i \partial_r \bar{Y}^i{}_b$
and $\bar{W}^a{}_{bc} \equiv \bar{X}^a{}_i \partial_r \bar{Y}^i{}_{bc}$.
The right-hand-side terms involving time derivatives
can be evaluated by using the other boundary conditions:
Differentiating the Sommerfeld condition~\eqref{eqn-Sommerfeld}
for~$\cphi$ gives
\begin{equation}
\partial_i \partial_t \cphi
	= -c \left[
	     \partial_{ir} \delta\cphi
	     + \frac{n}{r} \partial_i \delta\cphi
	     + \left\{
	       \begin{tabular}{c@{\quad}l}
	       $- n/r^2$	& if $i = r$	\\
	       0		& otherwise	
	       \end{tabular}
	       \right\}
	     \right]
\end{equation}
while $\partial_t \tilde{g}^{ab}$ can be evaluated by applying the
matrix identity
$d\tilde{g}^{ab} = - \tilde{g}^{ai} \tilde{g}^{bj} d\tilde{g}_{ij}$
to the boundary condition~\eqref{eqn-Sdd-outer-BC}.

For $\alpha$, $K$, and $\cphi$, I use a propagation speed~$c$ given by
the Bona-Masso lapse propagation speed~\eqref{eqn-Bona-Masso-lapse-speed},
while for $\tilde{g}_{ij}$, $\tilde{A}_{ij}$, and $\tilde{\Gamma}^i$
I use a propagation speed given by the outgoing light-cone speed
$\alpha/\!\sqrt{g_{rr}} - \beta^r$.  I use a falloff power of $n = 1$
for all the Sommerfeld boundary conditions.

For historical reasons, all the numerical results reported here were
obtained using a simple Dirichlet outer boundary
condition~$\partial_t \tilde{\Gamma}^i = 0$ instead of the
Sommerfeld condition~\eqref{eqn-Gamma-tilde-u-outer-BC}.
This probably contributes to the outer boundary instabilities I see,
and I plan to switch to the Sommerfeld condition for future computations.


\begin{acknowledgments}
I thank John Baker, Herbert Balasin, Robert Beig,
Bernd Br\"{u}gmann, Frank Herrmann, Denis Pollney, and Natascha Riahi
for assistance with deriving the BSSN equations
and the $\tilde{\Gamma}^i$ interpatch transformations.
I thank Miguel Alcubierre, Ian Hawke, Frank L\"{o}ffler, Denis Pollney,
Erik Schnetter, and Virginia Vitzthum
for valuable advice and assistance.
I thank an anonymous referee for a number of helpful comments.

I thank Goldie Rodgers,
Jeremy Thorn,
the Alexander von Humboldt Foundation,
the AEI visitors program,
and the AEI postdoctoral fellowship program
for financial support.
I thank Peter C.~Aichelburg
and the Universit\"{a}t Wien Institut f\"{u}r Theoretische Physik
for hospitality and use of research facilities
during 1998-2001.
I thank the AEI
for hospitality and use of research facilities
during my 1998-2000 AEI visits.
I thank Jean Wolfgang, Peter Luckham, and Ulrich Kiermayr
for assistance with computer facilities.
The numerical results presented in this paper were obtained on a
variety of computer systems, notably the \program{Peyote} cluster
at the AEI.
\end{acknowledgments}


\bibliography{references}

\begin{thebibliography}{80}
\expandafter\ifx\csname natexlab\endcsname\relax\def\natexlab#1{#1}\fi
\expandafter\ifx\csname bibnamefont\endcsname\relax
  \def\bibnamefont#1{#1}\fi
\expandafter\ifx\csname bibfnamefont\endcsname\relax
  \def\bibfnamefont#1{#1}\fi
\expandafter\ifx\csname citenamefont\endcsname\relax
  \def\citenamefont#1{#1}\fi
\expandafter\ifx\csname url\endcsname\relax
  \def\url#1{\texttt{#1}}\fi
\expandafter\ifx\csname urlprefix\endcsname\relax\def\urlprefix{URL }\fi
\providecommand{\bibinfo}[2]{#2}
\providecommand{\eprint}[2][]{\url{#2}}

\bibitem[{\citenamefont{Alcubierre}
  \emph{et~al.}(2004)\citenamefont{Alcubierre, Allen, Baumgarte, Bona, Fiske,
  Goodale, Guzm\'an, Hawke, Hawley, Husa, Koppitz, Lechner}
  \emph{et~al.}}]{Alcubierre2003:mexico-I}
\bibinfo{author}{\bibnamefont{Alcubierre}, \bibfnamefont{M.}},
  \bibinfo{author}{\bibfnamefont{G.}~\bibnamefont{Allen}},
  \bibinfo{author}{\bibfnamefont{T.~W.} \bibnamefont{Baumgarte}},
  \bibinfo{author}{\bibfnamefont{C.}~\bibnamefont{Bona}},
  \bibinfo{author}{\bibfnamefont{D.}~\bibnamefont{Fiske}},
  \bibinfo{author}{\bibfnamefont{T.}~\bibnamefont{Goodale}},
  \bibinfo{author}{\bibfnamefont{F.~S.} \bibnamefont{Guzm\'an}},
  \bibinfo{author}{\bibfnamefont{I.}~\bibnamefont{Hawke}},
  \bibinfo{author}{\bibfnamefont{S.}~\bibnamefont{Hawley}},
  \bibinfo{author}{\bibfnamefont{S.}~\bibnamefont{Husa}},
  \bibinfo{author}{\bibfnamefont{M.}~\bibnamefont{Koppitz}},
  \bibinfo{author}{\bibfnamefont{C.}~\bibnamefont{Lechner}}, \emph{et~al.},
  \bibinfo{year}{2004}, \bibinfo{journal}{Class. Quantum Grav.}
  \textbf{\bibinfo{volume}{21}}(\bibinfo{number}{2}), \bibinfo{pages}{589}.

\bibitem[{\citenamefont{Alcubierre}
  \emph{et~al.}(2003{\natexlab{a}})\citenamefont{Alcubierre, Br\"ugmann,
  Diener, Koppitz, Pollney, Seidel, and Takahashi}}]{Alcubierre02a}
\bibinfo{author}{\bibnamefont{Alcubierre}, \bibfnamefont{M.}},
  \bibinfo{author}{\bibfnamefont{B.}~\bibnamefont{Br\"ugmann}},
  \bibinfo{author}{\bibfnamefont{P.}~\bibnamefont{Diener}},
  \bibinfo{author}{\bibfnamefont{M.}~\bibnamefont{Koppitz}},
  \bibinfo{author}{\bibfnamefont{D.}~\bibnamefont{Pollney}},
  \bibinfo{author}{\bibfnamefont{E.}~\bibnamefont{Seidel}}, and
  \bibinfo{author}{\bibfnamefont{R.}~\bibnamefont{Takahashi}},
  \bibinfo{year}{2003}{\natexlab{a}}, \bibinfo{journal}{Phys. Rev. D}
  \textbf{\bibinfo{volume}{67}}, \bibinfo{pages}{084023}.

\bibitem[{\citenamefont{Alcubierre}
  \emph{et~al.}(2000)\citenamefont{Alcubierre, Br\"{u}gmann, Dramlitsch, Font,
  Papadopoulos, Seidel, Stergioulas, and Takahashi}}]{Alcubierre99d}
\bibinfo{author}{\bibnamefont{Alcubierre}, \bibfnamefont{M.}},
  \bibinfo{author}{\bibfnamefont{B.}~\bibnamefont{Br\"{u}gmann}},
  \bibinfo{author}{\bibfnamefont{T.}~\bibnamefont{Dramlitsch}},
  \bibinfo{author}{\bibfnamefont{J.}~\bibnamefont{Font}},
  \bibinfo{author}{\bibfnamefont{P.}~\bibnamefont{Papadopoulos}},
  \bibinfo{author}{\bibfnamefont{E.}~\bibnamefont{Seidel}},
  \bibinfo{author}{\bibfnamefont{N.}~\bibnamefont{Stergioulas}}, and
  \bibinfo{author}{\bibfnamefont{R.}~\bibnamefont{Takahashi}},
  \bibinfo{year}{2000}, \bibinfo{journal}{Phys. Rev. D}
  \textbf{\bibinfo{volume}{62}}, \bibinfo{pages}{044034}.

\bibitem[{\citenamefont{Alcubierre}
  \emph{et~al.}(2003{\natexlab{b}})\citenamefont{Alcubierre, Corichi,
  Gonz\'alez, {N\'u\~nez}, and Salgado}}]{Alcubierre2003:hyperbolic-slicing}
\bibinfo{author}{\bibnamefont{Alcubierre}, \bibfnamefont{M.}},
  \bibinfo{author}{\bibfnamefont{A.}~\bibnamefont{Corichi}},
  \bibinfo{author}{\bibfnamefont{J.~A.} \bibnamefont{Gonz\'alez}},
  \bibinfo{author}{\bibfnamefont{D.}~\bibnamefont{{N\'u\~nez}}}, and
  \bibinfo{author}{\bibfnamefont{M.}~\bibnamefont{Salgado}},
  \bibinfo{year}{2003}{\natexlab{b}}, \bibinfo{journal}{Classical and Quantum
  Gravity} \textbf{\bibinfo{volume}{20}}(\bibinfo{number}{18}),
  \bibinfo{pages}{3951}.

\bibitem[{\citenamefont{Alcubierre and Mass{\'o}}(1998)}]{Alcubierre97b}
\bibinfo{author}{\bibnamefont{Alcubierre}, \bibfnamefont{M.}}, and
  \bibinfo{author}{\bibfnamefont{J.}~\bibnamefont{Mass{\'o}}},
  \bibinfo{year}{1998}, \bibinfo{journal}{Phys. Rev. D}
  \textbf{\bibinfo{volume}{57}}(\bibinfo{number}{8}), \bibinfo{pages}{4511}.

\bibitem[{\citenamefont{Alcubierre and Schutz}(1996)}]{Alcubierre94b}
\bibinfo{author}{\bibnamefont{Alcubierre}, \bibfnamefont{M.}}, and
  \bibinfo{author}{\bibfnamefont{B.}~\bibnamefont{Schutz}},
  \bibinfo{year}{1996}, in \emph{\bibinfo{booktitle}{The Seventh {M}arcel
  {G}rossmann Meeting: On Recent Developments in Theoretical and Experimental
  General Relativity, Gravitation, and Relativistic Field Theories}}, edited by
  \bibinfo{editor}{\bibfnamefont{R.~T.} \bibnamefont{Jantzen}},
  \bibinfo{editor}{\bibfnamefont{G.~M.} \bibnamefont{Keiser}}, and
  \bibinfo{editor}{\bibfnamefont{R.}~\bibnamefont{Ruffini}}
  (\bibinfo{publisher}{World {S}cientific}, \bibinfo{address}{Singapore}), p.
  \bibinfo{pages}{611}.

\bibitem[{\citenamefont{Anninos} \emph{et~al.}(1995)\citenamefont{Anninos,
  Daues, Mass{\'o}, Seidel, and Suen}}]{Anninos94e}
\bibinfo{author}{\bibnamefont{Anninos}, \bibfnamefont{P.}},
  \bibinfo{author}{\bibfnamefont{G.}~\bibnamefont{Daues}},
  \bibinfo{author}{\bibfnamefont{J.}~\bibnamefont{Mass{\'o}}},
  \bibinfo{author}{\bibfnamefont{E.}~\bibnamefont{Seidel}}, and
  \bibinfo{author}{\bibfnamefont{W.-M.} \bibnamefont{Suen}},
  \bibinfo{year}{1995}, \bibinfo{journal}{Phys. Rev. D}
  \textbf{\bibinfo{volume}{51}}(\bibinfo{number}{10}), \bibinfo{pages}{5562}.

\bibitem[{\citenamefont{Ansorg} \emph{et~al.}(2004)\citenamefont{Ansorg,
  Fischer, Kleinw\"achter, Meinel, Petroff, and Sch\"obel}}]{Ansorg:2004vv}
\bibinfo{author}{\bibnamefont{Ansorg}, \bibfnamefont{M.}},
  \bibinfo{author}{\bibfnamefont{T.}~\bibnamefont{Fischer}},
  \bibinfo{author}{\bibfnamefont{A.}~\bibnamefont{Kleinw\"achter}},
  \bibinfo{author}{\bibfnamefont{R.}~\bibnamefont{Meinel}},
  \bibinfo{author}{\bibfnamefont{D.}~\bibnamefont{Petroff}}, and
  \bibinfo{author}{\bibfnamefont{K.}~\bibnamefont{Sch\"obel}},
  \bibinfo{year}{2004}, \eprint{gr-qc/0402102}.

\bibitem[{\citenamefont{Ansorg} \emph{et~al.}(2003)\citenamefont{Ansorg,
  Kleinw\"achter, and Meinel}}]{Ansorg:2003br}
\bibinfo{author}{\bibnamefont{Ansorg}, \bibfnamefont{M.}},
  \bibinfo{author}{\bibfnamefont{A.}~\bibnamefont{Kleinw\"achter}}, and
  \bibinfo{author}{\bibfnamefont{R.}~\bibnamefont{Meinel}},
  \bibinfo{year}{2003}, \bibinfo{journal}{Astron. Astrophys.}
  \textbf{\bibinfo{volume}{405}}, \bibinfo{pages}{711}.

\bibitem[{\citenamefont{Baumgarte and Shapiro}(1999)}]{Baumgarte99}
\bibinfo{author}{\bibnamefont{Baumgarte}, \bibfnamefont{T.~W.}}, and
  \bibinfo{author}{\bibfnamefont{S.~L.} \bibnamefont{Shapiro}},
  \bibinfo{year}{1999}, \bibinfo{journal}{Physical Review D}
  \textbf{\bibinfo{volume}{59}}, \bibinfo{pages}{024007}.

\bibitem[{\citenamefont{Berger}(1982)}]{Berger-1982}
\bibinfo{author}{\bibnamefont{Berger}, \bibfnamefont{M.~J.}},
  \bibinfo{year}{1982}, \emph{\bibinfo{title}{Adaptive Mesh Refinement for
  Hyperbolic Partial Differential Equations}}, Ph.D. thesis,
  \bibinfo{school}{Stanford University}, \bibinfo{note}{{U}niversity
  {M}icrofilms \#DA 83-01196}.

\bibitem[{\citenamefont{Berger}(1986)}]{Berger86}
\bibinfo{author}{\bibnamefont{Berger}, \bibfnamefont{M.~J.}},
  \bibinfo{year}{1986}, \bibinfo{journal}{SIAM Journal of Scientific and
  Statistical Computing} \textbf{\bibinfo{volume}{7}}(\bibinfo{number}{3}),
  \bibinfo{pages}{904}.

\bibitem[{\citenamefont{Berger and Colella}(1989)}]{Berger89}
\bibinfo{author}{\bibnamefont{Berger}, \bibfnamefont{M.~J.}}, and
  \bibinfo{author}{\bibfnamefont{P.}~\bibnamefont{Colella}},
  \bibinfo{year}{1989}, \bibinfo{journal}{J. Comp. Phys.}
  \textbf{\bibinfo{volume}{82}}, \bibinfo{pages}{64}.

\bibitem[{\citenamefont{Berger and Oliger}(1984)}]{Berger84}
\bibinfo{author}{\bibnamefont{Berger}, \bibfnamefont{M.~J.}}, and
  \bibinfo{author}{\bibfnamefont{J.}~\bibnamefont{Oliger}},
  \bibinfo{year}{1984}, \bibinfo{journal}{Journal of Computational Physics}
  \textbf{\bibinfo{volume}{53}}, \bibinfo{pages}{484}.

\bibitem[{\citenamefont{Bishop} \emph{et~al.}(1996)\citenamefont{Bishop,
  G\'{o}mez, Holvorcem, Matzner, Papadopoulos, and
  Winicour}}]{Bishop-etal-1996:Cauchy-characteristic-matching}
\bibinfo{author}{\bibnamefont{Bishop}, \bibfnamefont{N.~T.}},
  \bibinfo{author}{\bibfnamefont{R.}~\bibnamefont{G\'{o}mez}},
  \bibinfo{author}{\bibfnamefont{P.~R.} \bibnamefont{Holvorcem}},
  \bibinfo{author}{\bibfnamefont{R.~A.} \bibnamefont{Matzner}},
  \bibinfo{author}{\bibfnamefont{P.}~\bibnamefont{Papadopoulos}}, and
  \bibinfo{author}{\bibfnamefont{J.}~\bibnamefont{Winicour}},
  \bibinfo{year}{1996}, \bibinfo{journal}{Physical Review Letters}
  \textbf{\bibinfo{volume}{76}}(\bibinfo{number}{23}), \bibinfo{pages}{4303}.

\bibitem[{\citenamefont{Bishop} \emph{et~al.}(1997)\citenamefont{Bishop,
  G\'{o}mez, Lehner, Maharaj, and Winicour}}]{Bishop97b}
\bibinfo{author}{\bibnamefont{Bishop}, \bibfnamefont{N.~T.}},
  \bibinfo{author}{\bibfnamefont{R.}~\bibnamefont{G\'{o}mez}},
  \bibinfo{author}{\bibfnamefont{L.}~\bibnamefont{Lehner}},
  \bibinfo{author}{\bibfnamefont{M.}~\bibnamefont{Maharaj}}, and
  \bibinfo{author}{\bibfnamefont{J.}~\bibnamefont{Winicour}},
  \bibinfo{year}{1997}, \bibinfo{journal}{Phys. Rev. D}
  \textbf{\bibinfo{volume}{56}}(\bibinfo{number}{10}), \bibinfo{pages}{6298}.

\bibitem[{\citenamefont{Blum}(1962)}]{Blum-1962:low-storage-Runge-Kutta}
\bibinfo{author}{\bibnamefont{Blum}, \bibfnamefont{E.~K.}},
  \bibinfo{year}{1962}, \bibinfo{journal}{Mathematics of Computation}
  \textbf{\bibinfo{volume}{16}}(\bibinfo{number}{78}), \bibinfo{pages}{176}.

\bibitem[{\citenamefont{Bona} \emph{et~al.}(1995)\citenamefont{Bona, Mass{\'o},
  Seidel, and Stela}}]{Bona94b}
\bibinfo{author}{\bibnamefont{Bona}, \bibfnamefont{C.}},
  \bibinfo{author}{\bibfnamefont{J.}~\bibnamefont{Mass{\'o}}},
  \bibinfo{author}{\bibfnamefont{E.}~\bibnamefont{Seidel}}, and
  \bibinfo{author}{\bibfnamefont{J.}~\bibnamefont{Stela}},
  \bibinfo{year}{1995}, \bibinfo{journal}{Phys. Rev. Lett.}
  \textbf{\bibinfo{volume}{75}}, \bibinfo{pages}{600}.

\bibitem[{\citenamefont{Boyd}(2000)}]{Boyd00}
\bibinfo{author}{\bibnamefont{Boyd}, \bibfnamefont{J.~P.}},
  \bibinfo{year}{2000}, \emph{\bibinfo{title}{Chebyshev and Fourier Spectral
  Methods (Second Edition, Revised)}} (\bibinfo{publisher}{Dover Publications},
  \bibinfo{address}{New York}).

\bibitem[{\citenamefont{Brandt and Br{\"u}gmann}(1997)}]{Brandt97b}
\bibinfo{author}{\bibnamefont{Brandt}, \bibfnamefont{S.}}, and
  \bibinfo{author}{\bibfnamefont{B.}~\bibnamefont{Br{\"u}gmann}},
  \bibinfo{year}{1997}, \bibinfo{journal}{Phys. Rev. Lett.}
  \textbf{\bibinfo{volume}{78}}(\bibinfo{number}{19}), \bibinfo{pages}{3606}.

\bibitem[{\citenamefont{Brown} \emph{et~al.}(1997)\citenamefont{Brown,
  Chesshire, Henshaw, and Quinlan}}]{Brown-etal-1997:Overture}
\bibinfo{author}{\bibnamefont{Brown}, \bibfnamefont{D.~L.}},
  \bibinfo{author}{\bibfnamefont{G.~S.} \bibnamefont{Chesshire}},
  \bibinfo{author}{\bibfnamefont{W.~D.} \bibnamefont{Henshaw}}, and
  \bibinfo{author}{\bibfnamefont{D.~J.} \bibnamefont{Quinlan}},
  \bibinfo{year}{1997}, in \emph{\bibinfo{booktitle}{Eighth SIAM Conference on
  Parallel Processing for Scientific Computing, held in Minneapolis, Minnesota,
  14--17 March 1997}}, \urlprefix\url{http://www.llnl.gov/CASC/Overture/}.

\bibitem[{\citenamefont{Br\"ugmann}
  \emph{et~al.}(2004)\citenamefont{Br\"ugmann, Tichy, and
  Jansen}}]{Bruegmann:2003aw}
\bibinfo{author}{\bibnamefont{Br\"ugmann}, \bibfnamefont{B.}},
  \bibinfo{author}{\bibfnamefont{W.}~\bibnamefont{Tichy}}, and
  \bibinfo{author}{\bibfnamefont{N.}~\bibnamefont{Jansen}},
  \bibinfo{year}{2004}, \bibinfo{journal}{Phys. Rev. Lett.}
  \textbf{\bibinfo{volume}{92}}, \bibinfo{pages}{211101}.

\bibitem[{\citenamefont{Calabrese}
  \emph{et~al.}(2003{\natexlab{a}})\citenamefont{Calabrese, Lehner, Neilsen,
  Pullin, Reula, Sarbach, and Tiglio}}]{Calabrese2003a}
\bibinfo{author}{\bibnamefont{Calabrese}, \bibfnamefont{G.}},
  \bibinfo{author}{\bibfnamefont{L.}~\bibnamefont{Lehner}},
  \bibinfo{author}{\bibfnamefont{D.}~\bibnamefont{Neilsen}},
  \bibinfo{author}{\bibfnamefont{J.}~\bibnamefont{Pullin}},
  \bibinfo{author}{\bibfnamefont{O.}~\bibnamefont{Reula}},
  \bibinfo{author}{\bibfnamefont{O.}~\bibnamefont{Sarbach}}, and
  \bibinfo{author}{\bibfnamefont{M.}~\bibnamefont{Tiglio}},
  \bibinfo{year}{2003}{\natexlab{a}}, \bibinfo{journal}{Class. Quantum Grav}
  \textbf{\bibinfo{volume}{20}}, \bibinfo{pages}{L245}.

\bibitem[{\citenamefont{Calabrese}
  \emph{et~al.}(2003{\natexlab{b}})\citenamefont{Calabrese, Lehner, Reula,
  Sarbach, and Tiglio}}]{Calabrese:2003vx}
\bibinfo{author}{\bibnamefont{Calabrese}, \bibfnamefont{G.}},
  \bibinfo{author}{\bibfnamefont{L.}~\bibnamefont{Lehner}},
  \bibinfo{author}{\bibfnamefont{O.}~\bibnamefont{Reula}},
  \bibinfo{author}{\bibfnamefont{O.}~\bibnamefont{Sarbach}}, and
  \bibinfo{author}{\bibfnamefont{M.}~\bibnamefont{Tiglio}},
  \bibinfo{year}{2003}{\natexlab{b}}, \eprint{gr-qc/0308007}.

\bibitem[{\citenamefont{Calabrese and
  Neilsen}(2004{\natexlab{a}})}]{Calabrese-Neilsen-2004:multipatch-email-with-%
Thornburg}
\bibinfo{author}{\bibnamefont{Calabrese}, \bibfnamefont{G.}}, and
  \bibinfo{author}{\bibfnamefont{D.}~\bibnamefont{Neilsen}},
  \bibinfo{year}{2004}{\natexlab{a}}, \bibinfo{title}{personal communication to
  {J}onathan {T}hornburg}.

\bibitem[{\citenamefont{Calabrese and
  Neilsen}(2004{\natexlab{b}})}]{Calabrese2003:excision-and-summation-by-parts}
\bibinfo{author}{\bibnamefont{Calabrese}, \bibfnamefont{G.}}, and
  \bibinfo{author}{\bibfnamefont{D.}~\bibnamefont{Neilsen}},
  \bibinfo{year}{2004}{\natexlab{b}}, \bibinfo{journal}{Physical Review~D}
  \textbf{\bibinfo{volume}{69}}, \bibinfo{pages}{044020 (21~pages)}.

\bibitem[{\citenamefont{Carpenter} \emph{et~al.}(1999)\citenamefont{Carpenter,
  Nordstr\"{o}m, and
  Gottlieb}}]{Carpenter-etal-1999:high-order-multiple-patch-FD}
\bibinfo{author}{\bibnamefont{Carpenter}, \bibfnamefont{M.~H.}},
  \bibinfo{author}{\bibfnamefont{J.}~\bibnamefont{Nordstr\"{o}m}}, and
  \bibinfo{author}{\bibfnamefont{D.}~\bibnamefont{Gottlieb}},
  \bibinfo{year}{1999}, \bibinfo{journal}{Journal of Computational Physics}
  \textbf{\bibinfo{volume}{148}}(\bibinfo{number}{2}), \bibinfo{pages}{341}.

\bibitem[{\citenamefont{Chesshire and
  Henshaw}(1990)}]{Chesshire-Henshaw-1990:overlapping-grids-PDEs}
\bibinfo{author}{\bibnamefont{Chesshire}, \bibfnamefont{G.~S.}}, and
  \bibinfo{author}{\bibfnamefont{W.~D.} \bibnamefont{Henshaw}},
  \bibinfo{year}{1990}, \bibinfo{journal}{Journal of Computational Physics}
  \textbf{\bibinfo{volume}{90}}(\bibinfo{number}{1}), \bibinfo{pages}{1}.

\bibitem[{\citenamefont{Chesshire and
  Henshaw}(1994)}]{Chesshire-Henshaw-1994:overlapping-grids-conservative-inter%
polation}
\bibinfo{author}{\bibnamefont{Chesshire}, \bibfnamefont{G.~S.}}, and
  \bibinfo{author}{\bibfnamefont{W.~D.} \bibnamefont{Henshaw}},
  \bibinfo{year}{1994}, \bibinfo{journal}{SIAM Journal of Scientific Computing}
  \textbf{\bibinfo{volume}{15}}(\bibinfo{number}{4}), \bibinfo{pages}{819}.

\bibitem[{\citenamefont{Choptuik}(1986)}]{Choptuik86}
\bibinfo{author}{\bibnamefont{Choptuik}, \bibfnamefont{M.~W.}},
  \bibinfo{year}{1986}, \emph{\bibinfo{title}{A Study of Numerical Techniques
  for Radiative Problems in General Relativity}}, Ph.D. thesis,
  \bibinfo{school}{University of {B}ritish {C}olumbia}.

\bibitem[{\citenamefont{Choptuik}(1989)}]{Choptuik89}
\bibinfo{author}{\bibnamefont{Choptuik}, \bibfnamefont{M.~W.}},
  \bibinfo{year}{1989}, in \emph{\bibinfo{booktitle}{Frontiers in Numerical
  Relativity}}, edited by
  \bibinfo{editor}{\bibfnamefont{C.}~\bibnamefont{Evans}},
  \bibinfo{editor}{\bibfnamefont{L.}~\bibnamefont{Finn}}, and
  \bibinfo{editor}{\bibfnamefont{D.}~\bibnamefont{Hobill}}
  (\bibinfo{publisher}{Cambridge University Press},
  \bibinfo{address}{Cambridge, England}), pp. \bibinfo{pages}{206--221}.

\bibitem[{\citenamefont{Choptuik}(1991)}]{Choptuik91}
\bibinfo{author}{\bibnamefont{Choptuik}, \bibfnamefont{M.~W.}},
  \bibinfo{year}{1991}, \bibinfo{journal}{Phys. Rev. D}
  \textbf{\bibinfo{volume}{44}}, \bibinfo{pages}{3124}.

\bibitem[{\citenamefont{Gary}(1975)}]{Gary1975:MOL-outflow-BC}
\bibinfo{author}{\bibnamefont{Gary}, \bibfnamefont{J.}}, \bibinfo{year}{1975},
  \emph{\bibinfo{title}{Boundary Conditions for the Method of Lines Applied to
  Hyperbolic Systems}}, \bibinfo{type}{Technical Report}
  \bibinfo{number}{CU-CS-073-75}, \bibinfo{institution}{Department of Computer
  Science, University of Colorado}.

\bibitem[{\citenamefont{G\'{o}mez} \emph{et~al.}(1998)\citenamefont{G\'{o}mez,
  Lehner, Marsa, and Winicour}}]{Gomez97a}
\bibinfo{author}{\bibnamefont{G\'{o}mez}, \bibfnamefont{R.}},
  \bibinfo{author}{\bibfnamefont{L.}~\bibnamefont{Lehner}},
  \bibinfo{author}{\bibfnamefont{R.}~\bibnamefont{Marsa}}, and
  \bibinfo{author}{\bibfnamefont{J.}~\bibnamefont{Winicour}},
  \bibinfo{year}{1998}, \bibinfo{journal}{Phys. Rev. D}
  \textbf{\bibinfo{volume}{57}}(\bibinfo{number}{8}), \bibinfo{pages}{4778}.

\bibitem[{\citenamefont{G\'{o}mez} \emph{et~al.}(1997)\citenamefont{G\'{o}mez,
  Lehner, Papadopoulos, and Winicour}}]{Gomez97}
\bibinfo{author}{\bibnamefont{G\'{o}mez}, \bibfnamefont{R.}},
  \bibinfo{author}{\bibfnamefont{L.}~\bibnamefont{Lehner}},
  \bibinfo{author}{\bibfnamefont{P.}~\bibnamefont{Papadopoulos}}, and
  \bibinfo{author}{\bibfnamefont{J.}~\bibnamefont{Winicour}},
  \bibinfo{year}{1997}, \bibinfo{journal}{Classical and Quantum Gravity}
  \textbf{\bibinfo{volume}{14}}(\bibinfo{number}{4}), \bibinfo{pages}{977}.

\bibitem[{\citenamefont{Goodale} \emph{et~al.}(2003)\citenamefont{Goodale,
  Allen, Lanfermann, Mass\'o, Radke, Seidel, and Shalf}}]{Goodale02a}
\bibinfo{author}{\bibnamefont{Goodale}, \bibfnamefont{T.}},
  \bibinfo{author}{\bibfnamefont{G.}~\bibnamefont{Allen}},
  \bibinfo{author}{\bibfnamefont{G.}~\bibnamefont{Lanfermann}},
  \bibinfo{author}{\bibfnamefont{J.}~\bibnamefont{Mass\'o}},
  \bibinfo{author}{\bibfnamefont{T.}~\bibnamefont{Radke}},
  \bibinfo{author}{\bibfnamefont{E.}~\bibnamefont{Seidel}}, and
  \bibinfo{author}{\bibfnamefont{J.}~\bibnamefont{Shalf}},
  \bibinfo{year}{2003}, in \emph{\bibinfo{booktitle}{Vector and Parallel
  Processing - VECPAR'2002, 5th International Conference, Lecture Notes in
  Computer Science}} (\bibinfo{publisher}{Springer},
  \bibinfo{address}{Berlin}).

\bibitem[{\citenamefont{Grandcl\'{e}ment}
  \emph{et~al.}(2001)\citenamefont{Grandcl\'{e}ment, Bonazzola, Gourgoulhon,
  and Marck}}]{Grandclement-etal-2000:multi-domain-spectral-method}
\bibinfo{author}{\bibnamefont{Grandcl\'{e}ment}, \bibfnamefont{P.}},
  \bibinfo{author}{\bibfnamefont{S.}~\bibnamefont{Bonazzola}},
  \bibinfo{author}{\bibfnamefont{E.}~\bibnamefont{Gourgoulhon}}, and
  \bibinfo{author}{\bibfnamefont{J.-A.} \bibnamefont{Marck}},
  \bibinfo{year}{2001}, \bibinfo{journal}{Journal of Computational Physics}
  \textbf{\bibinfo{volume}{170}}, \bibinfo{pages}{231}.

\bibitem[{\citenamefont{Grandcl\'{e}ment}
  \emph{et~al.}(2002)\citenamefont{Grandcl\'{e}ment, Gourgoulhon, and
  Bonazzola}}]{Grandclement02}
\bibinfo{author}{\bibnamefont{Grandcl\'{e}ment}, \bibfnamefont{P.}},
  \bibinfo{author}{\bibfnamefont{E.}~\bibnamefont{Gourgoulhon}}, and
  \bibinfo{author}{\bibfnamefont{S.}~\bibnamefont{Bonazzola}},
  \bibinfo{year}{2002}, \bibinfo{journal}{Phys. Rev. D}
  \textbf{\bibinfo{volume}{65}}, \bibinfo{pages}{044021}.

\bibitem[{\citenamefont{Gustafsson}(1971)}]{Gustafsson1971:hyperbolic-BC-FD-co%
nvergence}
\bibinfo{author}{\bibnamefont{Gustafsson}, \bibfnamefont{B.}},
  \bibinfo{year}{1971}, \emph{\bibinfo{title}{On the Convergence Rate for
  Difference Approximations to Mixed Initial Boundary Value Problems}},
  \bibinfo{type}{Technical Report} \bibinfo{number}{33},
  \bibinfo{institution}{Department of Computer Science, Uppsala University}.

\bibitem[{\citenamefont{Gustafsson}(1975)}]{Gustafsson1975:hyperbolic-BC-FD-co%
nvergence}
\bibinfo{author}{\bibnamefont{Gustafsson}, \bibfnamefont{B.}},
  \bibinfo{year}{1975}, \bibinfo{journal}{Mathematics of Computation}
  \textbf{\bibinfo{volume}{29}}(\bibinfo{number}{130}), \bibinfo{pages}{396}.

\bibitem[{\citenamefont{Gustafsson}(1982)}]{Gustafsson1982:hyperbolic-BC-FD-co%
nvergence}
\bibinfo{author}{\bibnamefont{Gustafsson}, \bibfnamefont{B.}},
  \bibinfo{year}{1982}, \bibinfo{journal}{Journal of Computational Physics}
  \textbf{\bibinfo{volume}{48}}, \bibinfo{pages}{270}.

\bibitem[{\citenamefont{Gustafsson and Kreiss}(1979)}]{Gustafsson79}
\bibinfo{author}{\bibnamefont{Gustafsson}, \bibfnamefont{B.}}, and
  \bibinfo{author}{\bibfnamefont{H.-O.} \bibnamefont{Kreiss}},
  \bibinfo{year}{1979}, \bibinfo{journal}{J. Comp. Phys.}
  \textbf{\bibinfo{volume}{30}}, \bibinfo{pages}{333}.

\bibitem[{\citenamefont{Gustafsson}
  \emph{et~al.}(1995)\citenamefont{Gustafsson, Kreiss, and
  Oliger}}]{Gustafsson95}
\bibinfo{author}{\bibnamefont{Gustafsson}, \bibfnamefont{B.}},
  \bibinfo{author}{\bibfnamefont{H.-O.} \bibnamefont{Kreiss}}, and
  \bibinfo{author}{\bibfnamefont{J.}~\bibnamefont{Oliger}},
  \bibinfo{year}{1995}, \emph{\bibinfo{title}{Time dependent problems and
  difference methods}} (\bibinfo{publisher}{Wiley}, \bibinfo{address}{New
  York}).

\bibitem[{\citenamefont{Imbiriba} \emph{et~al.}(2004)\citenamefont{Imbiriba,
  Baker, Choi, Centrella, Fiske, Brown, van Meter, and
  Olson}}]{Imbiriba-etal-2004:puncture-evolution-FMR}
\bibinfo{author}{\bibnamefont{Imbiriba}, \bibfnamefont{B.}},
  \bibinfo{author}{\bibfnamefont{J.}~\bibnamefont{Baker}},
  \bibinfo{author}{\bibfnamefont{D.-I.} \bibnamefont{Choi}},
  \bibinfo{author}{\bibfnamefont{J.}~\bibnamefont{Centrella}},
  \bibinfo{author}{\bibfnamefont{D.~R.} \bibnamefont{Fiske}},
  \bibinfo{author}{\bibfnamefont{J.~D.} \bibnamefont{Brown}},
  \bibinfo{author}{\bibfnamefont{J.~R.} \bibnamefont{van Meter}}, and
  \bibinfo{author}{\bibfnamefont{K.}~\bibnamefont{Olson}},
  \bibinfo{year}{2004}, \bibinfo{title}{Evolving a puncture black hole with
  fixed mesh refinement}, \eprint{gr-qc/0403048}.

\bibitem[{\citenamefont{Kidder and Finn}(2000)}]{Kidder99a}
\bibinfo{author}{\bibnamefont{Kidder}, \bibfnamefont{L.~E.}}, and
  \bibinfo{author}{\bibfnamefont{L.~S.} \bibnamefont{Finn}},
  \bibinfo{year}{2000}, \bibinfo{journal}{Phys. Rev. D}
  \textbf{\bibinfo{volume}{62}}, \bibinfo{pages}{084026},
  \bibinfo{note}{gr-qc/9911014}.

\bibitem[{\citenamefont{Lehner}(2003)}]{Lehner-2003:Kerr-cubical-excision-prob%
lems}
\bibinfo{author}{\bibnamefont{Lehner}, \bibfnamefont{L.}},
  \bibinfo{year}{2003}, \bibinfo{title}{personal communication},
  \bibinfo{note}{cited in
  \citet{Calabrese2003:excision-and-summation-by-parts}}.

\bibitem[{\citenamefont{Lehner}
  \emph{et~al.}(2004{\natexlab{a}})\citenamefont{Lehner, Neilsen, Reula, and
  Tiglio}}]{Lehner-etal-2004:Kerr-cubical-excision-problems}
\bibinfo{author}{\bibnamefont{Lehner}, \bibfnamefont{L.}},
  \bibinfo{author}{\bibfnamefont{D.}~\bibnamefont{Neilsen}},
  \bibinfo{author}{\bibfnamefont{O.}~\bibnamefont{Reula}}, and
  \bibinfo{author}{\bibfnamefont{M.}~\bibnamefont{Tiglio}},
  \bibinfo{year}{2004}{\natexlab{a}}, \bibinfo{title}{paper in preparation}.

\bibitem[{\citenamefont{Lehner}
  \emph{et~al.}(2004{\natexlab{b}})\citenamefont{Lehner, Reula, and
  Tiglio}}]{Lehner-Reula-Tiglio-2004:multipatch-scalar-field-Kerr-background}
\bibinfo{author}{\bibnamefont{Lehner}, \bibfnamefont{L.}},
  \bibinfo{author}{\bibfnamefont{O.}~\bibnamefont{Reula}}, and
  \bibinfo{author}{\bibfnamefont{M.}~\bibnamefont{Tiglio}},
  \bibinfo{year}{2004}{\natexlab{b}}, \bibinfo{title}{paper in preparation}.

\bibitem[{\citenamefont{Lichnerowicz}(1944)}]{Lichnerowicz44}
\bibinfo{author}{\bibnamefont{Lichnerowicz}, \bibfnamefont{A.}},
  \bibinfo{year}{1944}, \bibinfo{journal}{J. Math Pures et Appl.}
  \textbf{\bibinfo{volume}{23}}, \bibinfo{pages}{37}.

\bibitem[{\citenamefont{Liebling}(2002)}]{Liebling-2002:nonlinear-sigma-critic%
ality-via-3D-AMR}
\bibinfo{author}{\bibnamefont{Liebling}, \bibfnamefont{S.~L.}},
  \bibinfo{year}{2002}, \bibinfo{journal}{Physical Review D}
  \textbf{\bibinfo{volume}{66}}, \bibinfo{pages}{041703}.

\bibitem[{\citenamefont{Misner} \emph{et~al.}(1973)\citenamefont{Misner,
  Thorne, and Wheeler}}]{Misner73}
\bibinfo{author}{\bibnamefont{Misner}, \bibfnamefont{C.~W.}},
  \bibinfo{author}{\bibfnamefont{K.~S.} \bibnamefont{Thorne}}, and
  \bibinfo{author}{\bibfnamefont{J.~A.} \bibnamefont{Wheeler}},
  \bibinfo{year}{1973}, \emph{\bibinfo{title}{Gravitation}}
  (\bibinfo{publisher}{W. H. Freeman}, \bibinfo{address}{San Francisco}).

\bibitem[{\citenamefont{Nakamura and Oohara}(1989)}]{Nakamura89}
\bibinfo{author}{\bibnamefont{Nakamura}, \bibfnamefont{T.}}, and
  \bibinfo{author}{\bibfnamefont{K.}~\bibnamefont{Oohara}},
  \bibinfo{year}{1989}, in \emph{\bibinfo{booktitle}{Frontiers in Numerical
  Relativity}}, edited by
  \bibinfo{editor}{\bibfnamefont{C.}~\bibnamefont{Evans}},
  \bibinfo{editor}{\bibfnamefont{L.}~\bibnamefont{Finn}}, and
  \bibinfo{editor}{\bibfnamefont{D.}~\bibnamefont{Hobill}}
  (\bibinfo{publisher}{Cambridge University Press},
  \bibinfo{address}{Cambridge, England}), pp. \bibinfo{pages}{254--280}.

\bibitem[{\citenamefont{Nakamura and Oohara}(1998)}]{Nakamura99a}
\bibinfo{author}{\bibnamefont{Nakamura}, \bibfnamefont{T.}}, and
  \bibinfo{author}{\bibfnamefont{K.}~\bibnamefont{Oohara}},
  \bibinfo{year}{1998}, in \emph{\bibinfo{booktitle}{Numerical Astrophysics
  1998 (NAP98) - Proceedings}}, \eprint{gr-qc/9812054},
  \urlprefix\url{http://www.a.phys.nagoya-u.ac.jp/NAP98-proceedings-official.h%
tml}.

\bibitem[{\citenamefont{Nakamura} \emph{et~al.}(1987)\citenamefont{Nakamura,
  Oohara, and Kojima}}]{Nakamura87}
\bibinfo{author}{\bibnamefont{Nakamura}, \bibfnamefont{T.}},
  \bibinfo{author}{\bibfnamefont{K.}~\bibnamefont{Oohara}}, and
  \bibinfo{author}{\bibfnamefont{Y.}~\bibnamefont{Kojima}},
  \bibinfo{year}{1987}, \bibinfo{journal}{Progress of Theoretical Physics
  Supplement} \textbf{\bibinfo{volume}{90}}, \bibinfo{pages}{1}.

\bibitem[{\citenamefont{Olsson and
  Petersson}(1996)}]{Olsson-Petersson-1996:overlapping-grid-stability}
\bibinfo{author}{\bibnamefont{Olsson}, \bibfnamefont{F.}}, and
  \bibinfo{author}{\bibfnamefont{N.~A.} \bibnamefont{Petersson}},
  \bibinfo{year}{1996}, \bibinfo{journal}{Computers and Fluids}
  \textbf{\bibinfo{volume}{25}}(\bibinfo{number}{6}), \bibinfo{pages}{583}.

\bibitem[{\citenamefont{Petersson}(1999)}]{Petersson-1999:overlapping-grid-gen%
eration}
\bibinfo{author}{\bibnamefont{Petersson}, \bibfnamefont{N.~A.}},
  \bibinfo{year}{1999}, \bibinfo{journal}{SIAM Journal of Scientific Computing}
  \textbf{\bibinfo{volume}{20}}, \bibinfo{pages}{1995},
  \urlprefix\url{http://www.llnl.gov/CASC/Overture/henshaw/publications/Peters%
sonAlgorithm.pdf}.

\bibitem[{\citenamefont{Pfeiffer} \emph{et~al.}(2003)\citenamefont{Pfeiffer,
  Kidder, Scheel, and Teukolsky}}]{Pfeiffer:2002wt}
\bibinfo{author}{\bibnamefont{Pfeiffer}, \bibfnamefont{H.~P.}},
  \bibinfo{author}{\bibfnamefont{L.~E.} \bibnamefont{Kidder}},
  \bibinfo{author}{\bibfnamefont{M.~A.} \bibnamefont{Scheel}}, and
  \bibinfo{author}{\bibfnamefont{S.~A.} \bibnamefont{Teukolsky}},
  \bibinfo{year}{2003}, \bibinfo{journal}{Comput. Phys. Commun.}
  \textbf{\bibinfo{volume}{152}}, \bibinfo{pages}{253}.

\bibitem[{\citenamefont{Reula}(2003)}]{Reula-2003:Trieste-talk}
\bibinfo{author}{\bibnamefont{Reula}, \bibfnamefont{O.}}, \bibinfo{year}{2003},
  \bibinfo{title}{Novel finite-differencing techniques for numerical
  relativity: Application to black-hole excision}, \bibinfo{note}{talk at the
  Advanced School \& Conference on Sources of Gravitational Waves, held at the
  Abdus Salam International Centre for Theoretical Physics, Trieste, Italy},
  \urlprefix\url{http://surubi.fis.uncor.edu/~reula/Seminars/ICTP/novel_f.pdf}.

\bibitem[{\citenamefont{Rubbert and
  Lee}(1982)}]{Rubbert-Lee-in-Thompson-1982:numerical-grid-generation}
\bibinfo{author}{\bibnamefont{Rubbert}, \bibfnamefont{P.~E.}}, and
  \bibinfo{author}{\bibfnamefont{K.~D.} \bibnamefont{Lee}},
  \bibinfo{year}{1982}, in \emph{\bibinfo{booktitle}{Numerical Grid
  Generation}}, edited by \bibinfo{editor}{\bibfnamefont{J.~F.}
  \bibnamefont{Thompson}} (\bibinfo{publisher}{North-Holland},
  \bibinfo{address}{New York}), pp. \bibinfo{pages}{235--252}, ISBN
  \bibinfo{isbn}{0-444-00757-1}.

\bibitem[{\citenamefont{Scheel and
  Kidder}(2000)}]{Scheel2000:Santa-Barbara-talk}
\bibinfo{author}{\bibnamefont{Scheel}, \bibfnamefont{M.~A.}}, and
  \bibinfo{author}{\bibfnamefont{L.}~\bibnamefont{Kidder}},
  \bibinfo{year}{2000}, \bibinfo{title}{Cornell/{NCSA}--$3{+}1$ {G}roup
  {R}eport}, \bibinfo{note}{talk at Miniprogram on Colliding Black Holes, held
  at the {I}nstitute for {T}heoretical {P}hysics, {U}niversity of {C}alifornia
  at {S}anta {B}arbara, 10-28 January 2000},
  \urlprefix\url{http://online.kitp.ucsb.edu/online/numrel00/}.

\bibitem[{\citenamefont{Schnetter} \emph{et~al.}(2004)\citenamefont{Schnetter,
  Hawley, and Hawke}}]{Schnetter-etal-03b}
\bibinfo{author}{\bibnamefont{Schnetter}, \bibfnamefont{E.}},
  \bibinfo{author}{\bibfnamefont{S.~H.} \bibnamefont{Hawley}}, and
  \bibinfo{author}{\bibfnamefont{I.}~\bibnamefont{Hawke}},
  \bibinfo{year}{2004}, \bibinfo{journal}{Class. Quantum Grav.}
  \textbf{\bibinfo{volume}{21}}(\bibinfo{number}{6}), \bibinfo{pages}{1465}.

\bibitem[{\citenamefont{Seidel and Suen}(1992)}]{Seidel92a}
\bibinfo{author}{\bibnamefont{Seidel}, \bibfnamefont{E.}}, and
  \bibinfo{author}{\bibfnamefont{W.-M.} \bibnamefont{Suen}},
  \bibinfo{year}{1992}, \bibinfo{journal}{Phys. Rev. Lett.}
  \textbf{\bibinfo{volume}{69}}(\bibinfo{number}{13}), \bibinfo{pages}{1845}.

\bibitem[{\citenamefont{Shibata}(1999)}]{Shibata99a}
\bibinfo{author}{\bibnamefont{Shibata}, \bibfnamefont{M.}},
  \bibinfo{year}{1999}, \bibinfo{journal}{Prog. Theor. Phys.}
  \textbf{\bibinfo{volume}{101}}, \bibinfo{pages}{1199},
  \bibinfo{note}{gr-qc/9905058}.

\bibitem[{\citenamefont{Shibata and Nakamura}(1995)}]{Shibata95}
\bibinfo{author}{\bibnamefont{Shibata}, \bibfnamefont{M.}}, and
  \bibinfo{author}{\bibfnamefont{T.}~\bibnamefont{Nakamura}},
  \bibinfo{year}{1995}, \bibinfo{journal}{Phys. Rev. D}
  \textbf{\bibinfo{volume}{52}}, \bibinfo{pages}{5428}.

\bibitem[{\citenamefont{Smarr and York}(1978)}]{Smarr78b}
\bibinfo{author}{\bibnamefont{Smarr}, \bibfnamefont{L.}}, and
  \bibinfo{author}{\bibfnamefont{J.~W.} \bibnamefont{York},
  \bibfnamefont{Jr.}}, \bibinfo{year}{1978}, \bibinfo{journal}{Phys. Rev. D}
  \textbf{\bibinfo{volume}{17}}(\bibinfo{number}{10}), \bibinfo{pages}{2529}.

\bibitem[{\citenamefont{Starius}(1980)}]{Starius-1980:composite-mesh-FD-for-hy%
perbolic-PDE}
\bibinfo{author}{\bibnamefont{Starius}, \bibfnamefont{G.}},
  \bibinfo{year}{1980}, \bibinfo{journal}{Numerische Mathematik}
  \textbf{\bibinfo{volume}{35}}, \bibinfo{pages}{241}.

\bibitem[{\citenamefont{Thompson} \emph{et~al.}(1985)\citenamefont{Thompson,
  Warsi, and Mastin}}]{Thompson-Warsi-Mastin-1985:numerical-grid-generation}
\bibinfo{author}{\bibnamefont{Thompson}, \bibfnamefont{J.~F.}},
  \bibinfo{author}{\bibfnamefont{Z.~U.~A.} \bibnamefont{Warsi}}, and
  \bibinfo{author}{\bibfnamefont{C.}~\bibnamefont{Mastin}},
  \bibinfo{year}{1985}, \emph{\bibinfo{title}{Numerical Grid Generation:
  Foundations and Applications}} (\bibinfo{publisher}{North-Holland},
  \bibinfo{address}{New York}), ISBN \bibinfo{isbn}{0-444-00985-X}.

\bibitem[{\citenamefont{Thornburg}(1985)}]{Thornburg85}
\bibinfo{author}{\bibnamefont{Thornburg}, \bibfnamefont{J.}},
  \bibinfo{year}{1985}, \emph{\bibinfo{title}{Coordinates and Boundary
  Conditions for the General Relativistic Initial Data Problem}}, Master's
  thesis, \bibinfo{school}{University of {B}ritish {C}olumbia},
  \bibinfo{address}{{V}ancouver, {B}ritish {C}olumbia}.

\bibitem[{\citenamefont{Thornburg}(1987)}]{Thornburg87}
\bibinfo{author}{\bibnamefont{Thornburg}, \bibfnamefont{J.}},
  \bibinfo{year}{1987}, \bibinfo{journal}{Class. Quantum Grav.}
  \textbf{\bibinfo{volume}{4}}, \bibinfo{pages}{1119}.

\bibitem[{\citenamefont{Thornburg}(1993)}]{Thornburg93}
\bibinfo{author}{\bibnamefont{Thornburg}, \bibfnamefont{J.}},
  \bibinfo{year}{1993}, \emph{\bibinfo{title}{Numerical Relativity in Black
  Hole Spacetimes}}, Ph.D. thesis, \bibinfo{school}{University of {B}ritish
  {C}olumbia}, \bibinfo{address}{{V}ancouver, {B}ritish {C}olumbia}.

\bibitem[{\citenamefont{Thornburg}(1999{\natexlab{a}})}]{Thornburg98}
\bibinfo{author}{\bibnamefont{Thornburg}, \bibfnamefont{J.}},
  \bibinfo{year}{1999}{\natexlab{a}}, \bibinfo{journal}{Phys. Rev. D}
  \textbf{\bibinfo{volume}{59}}, \bibinfo{pages}{104007 (26~pages)}.

\bibitem[{\citenamefont{Thornburg}(1999{\natexlab{b}})}]{Thornburg99}
\bibinfo{author}{\bibnamefont{Thornburg}, \bibfnamefont{J.}},
  \bibinfo{year}{1999}{\natexlab{b}}, \bibinfo{note}{gr-qc/9906022}.

\bibitem[{\citenamefont{Thornburg}(2003)}]{Thornburg2000:multiple-patch-evolut%
ion}
\bibinfo{author}{\bibnamefont{Thornburg}, \bibfnamefont{J.}},
  \bibinfo{year}{2003}, in \emph{\bibinfo{booktitle}{The Ninth {M}arcel
  {G}rossman Meeting: On Recent Developments in Theoretical and Experimental
  General Relavtivity, Gravitation, and Relativistic Field Theories}}, edited
  by \bibinfo{editor}{\bibfnamefont{V.~G.} \bibnamefont{Gurzadyan}},
  \bibinfo{editor}{\bibfnamefont{R.~T.} \bibnamefont{Jantzen}}, and
  \bibinfo{editor}{\bibfnamefont{R.}~\bibnamefont{Ruffini}}
  (\bibinfo{publisher}{World {S}cientific}, \bibinfo{address}{Singapore}), pp.
  \bibinfo{pages}{1743--1744}, \eprint{gr-qc/0012012}.

\bibitem[{\citenamefont{Thornburg}(2004)}]{Thornburg2003:AH-finding}
\bibinfo{author}{\bibnamefont{Thornburg}, \bibfnamefont{J.}},
  \bibinfo{year}{2004}, \bibinfo{journal}{Class. Quantum Grav.}
  \textbf{\bibinfo{volume}{21}}(\bibinfo{number}{2}), \bibinfo{pages}{743},
  \urlprefix\url{http://stacks.iop.org/0264-9381/21/743}.

\bibitem[{\citenamefont{Tiglio}(2003)}]{Tiglio-2003:Penn-State-talk}
\bibinfo{author}{\bibnamefont{Tiglio}, \bibfnamefont{M.}},
  \bibinfo{year}{2003}, \bibinfo{title}{Black hole evolutions: numerical
  techniques, boundary conditions and constraint preservation},
  \bibinfo{note}{talk at ``{G}ravitation: {A} {D}ecennial {P}erspective'', a
  conference held at the {P}enn {S}tate {C}enter for {G}ravitational {P}hysics,
  8-12 June 2003},
  \urlprefix\url{http://cgpg.gravity.psu.edu/events/conferences/Gravitation_De%
cennial/Proceedings/NR/Tuesday/Tiglio/tiglio_talk.pdf}.

\bibitem[{\citenamefont{Tiglio}(2004)}]{Tiglio-2004:multipatch-email-with-Thor%
nburg}
\bibinfo{author}{\bibnamefont{Tiglio}, \bibfnamefont{M.}},
  \bibinfo{year}{2004}, \bibinfo{title}{personal communication to {J}onathan
  {T}hornburg}.

\bibitem[{\citenamefont{Unruh}(1984)}]{Unruh84}
\bibinfo{author}{\bibnamefont{Unruh}, \bibfnamefont{W.~G.}},
  \bibinfo{year}{1984}, \bibinfo{title}{personal communication to {J}onathan
  {T}hornburg}.

\bibitem[{\citenamefont{Wald}(1984)}]{Wald84}
\bibinfo{author}{\bibnamefont{Wald}, \bibfnamefont{R.~M.}},
  \bibinfo{year}{1984}, \emph{\bibinfo{title}{General Relativity}}
  (\bibinfo{publisher}{The University of Chicago Press},
  \bibinfo{address}{Chicago}), ISBN \bibinfo{isbn}{0-226-87032-4 (hardcover),
  0-226-87033-2 (paperback)}.

\bibitem[{\citenamefont{Williamson}(1980)}]{Williamson-1980:low-storage-Runge-%
Kutta}
\bibinfo{author}{\bibnamefont{Williamson}, \bibfnamefont{J.~H.}},
  \bibinfo{year}{1980}, \bibinfo{journal}{Journal of Computational Physics}
  \textbf{\bibinfo{volume}{35}}(\bibinfo{number}{1}), \bibinfo{pages}{48}.

\bibitem[{\citenamefont{York}(1979)}]{York79}
\bibinfo{author}{\bibnamefont{York}, \bibfnamefont{J.~W., Jr.}},
  \bibinfo{year}{1979}, in \emph{\bibinfo{booktitle}{Sources of Gravitational
  Radiation}}, edited by \bibinfo{editor}{\bibfnamefont{L.~L.}
  \bibnamefont{Smarr}} (\bibinfo{publisher}{Cambridge University Press},
  \bibinfo{address}{Cambridge, UK}), ISBN \bibinfo{isbn}{0-521-22778-X}, pp.
  \bibinfo{pages}{83--126}.

\end{thebibliography}


\end{document}